\documentclass[a4paper,11pt]{article}
\usepackage{amssymb, amsmath}
\usepackage[active]{srcltx}

\newtheorem{theorem}{Theorem}[section]
\newtheorem{corollary}[theorem]{Corollary}
\newtheorem{lemma}[theorem]{Lemma}
\newtheorem{proposition}[theorem]{Proposition}
\newtheorem{definition}[theorem]{Definition}
\newtheorem{remark}[theorem]{Remark}
\numberwithin{equation}{section}

\title{Gibbs measures of disordered lattice systems with unbounded spins}
\author{Yuri Kondratiev$^a$, Yuri Kozitsky$^b$, and Tanja Pasurek$^a$\\[.3cm] \small{({\it a}) Fakult\"at f\"ur Mathematik, Universit\"at Bielefeld,}\\[.1cm] \small{ D-33615 Bielefeld, Germany}\\[.2cm] \small{({\it b}) Instytut Matematyki, Uniwersytet Marii Curie-Sk{\l}odowskiej},\\[.1cm] \small{20-031 Lublin, Poland} }

\begin{document}

\maketitle

\begin{abstract}
The Gibbs measures of a spin system on $\mathbb{Z}^d$ with unbounded pair interactions $J_{xy} \sigma(x) \sigma (y)$
are studied. Here $\langle x, y \rangle \in {\sf E}$, i.e. $x$ and $y$ are neighbors in $\mathbb{Z}^d$. 
The intensities $J_{xy}$ and the spins  $\sigma (x) , \sigma (y)$ are arbitrary real. To control their growth we introduce appropriate sets  $\mathcal{J}_q\subset \mathbb{R}^{\sf E}$ and $\mathcal{S}_p\subset \mathbb{R}^{\mathbb{Z}^d}$ and prove that  for every $J = (J_{xy}) \in \mathcal{J}_q$: (a) the set of Gibbs measures $\mathcal{G}_p(J)= \{\mu: {\rm solves} \ {\rm DLR}, \ \mu(\mathcal{S}_p)=1\}$ is non-void and weakly compact; (b) each $\mu\in\mathcal{G}_p(J)$ obeys an integrability estimate, the same for all $\mu$. Next we study the case where  $\mathcal{J}_q$ is equipped with a norm, with the Borel $\sigma$-field $\mathcal{B}(\mathcal{J}_q)$, and with a complete probability
measure $\nu$. We show that the set-valued map $\mathcal{J}_q \ni J \mapsto \mathcal{G}_p(J)$ is measurable and hence there exist measurable selections  $\mathcal{J}_q \ni J \mapsto \mu(J) \in \mathcal{G}_p(J)$, which are random Gibbs measures. We prove that the empirical distributions $N^{-1} \sum_{n=1}^N \pi_{\mathit{\Delta}_n} (\cdot| J, \xi)$, obtained from the local conditional Gibbs measures $\pi_{\mathit{\Delta}_n} (\cdot| J, \xi)$ and from exhausting sequences of $\mathit{\Delta}_n \subset \mathbb{Z}^d$, have $\nu$-a.s. weak limits as $N\rightarrow +\infty$, which are random Gibbs measures. Similarly, we prove the existence of the $\nu$-a.s. weak limits of the empirical metastates
$N^{-1} \sum_{n=1}^N \delta_{\pi_{\mathit{\Delta}_n} (\cdot| J,\xi)}$, which are Aizenman-Wehr metastates. Finally, we prove the existence of the limiting thermodynamic pressure
under some further conditions on $\nu$. The proof is based on a version of the first GKS inequality, which we obtain for our model.
\end{abstract}

\section{Introduction}

Throughout the paper, for a topological space, say $\mathcal{S}$, by $\mathcal{P}(\mathcal{S})$ we denote the set of all probability measures on $(\mathcal{S},\mathcal{B}(\mathcal{S}))$, where $\mathcal{B}(\mathcal{S})$ will always stand for the corresponding Borel $\sigma$-field.

Given  a countable set ${\sf X}$, a {\it random field} on ${\sf X}$  is a collection of random variables - {\it spins}, defined on some probability space and taking values in the corresponding {\it single-spin} (Polish) spaces $S_x$, $x\in {\sf X}$. In a `canonical
version', the probability space is $(\mathcal{S}, \mathcal{B}(\mathcal{S}), \mu)$, where $\mathcal{S}$ is the product space of all $S_x$. Then the notion {\it random field} is attributed to the latter measure as well. A particular case of such a field is the product measure of some {\it single-spin} probability measures $\chi_x$, $x\in {\sf X}$.
{\it Gibbs} random fields with {\it pair interactions} are constructed as perturbations of $\prod_{x\in {\sf X}}\chi_x$ by the `densities'
\begin{equation}
 \label{O}
 \exp\left( \sum_{\langle x,y \rangle} W_{xy} (\sigma (x) , \sigma (y)) \right),
\end{equation}
where $W_{xy} : S_x \times S_y \rightarrow \mathbb{R}$ are measurable functions -- {\it interaction potentials}, whereas the sum is taken over a subset of ${\sf X}\times {\sf X}$. Such a
field  defines the graph ${\sf G} = ({\sf X}, {\sf E})$, where the set of edges ${\sf E}$ consists of those pairs $\{ x, y\}$ where
$W_{xy}$ is not the zero function.  The case of a special interest is where the potentials are {\it random}. Then one deals with another random field, this time on ${\sf E}$, represented by the triple $(\mathcal{W}, \mathcal{F}, P)$. Here $\mathcal{W}$ is the space of interactions consisting of $W=(W_{xy})_{\langle x,y \rangle \in {\sf E}}$, $\mathcal{F}$ is an appropriate $\sigma$-field, and $P$ is a probability measure. A standard assumption is that the degree of each vertex is finite and that the functions
$W_{xy} : S_x \times S_y \rightarrow \mathbb{R}$ are $P$-almost surely bounded, in which case the interactions are called {\it regular}, c.f. Definition 6.2.1 in \cite{[Bovier]}, page 99. The only {\it irregular} case studied in the literature is that  of long-range spin glasses, where the single-spin spaces are finite, and thus the functions $W_{xy}$ are bounded, but the vertex degrees are infinite. In the case of regular $W$, a measurable map
\begin{equation} \label{Mes}
 \mathcal{W} \ni W \mapsto \mu(W) \in \mathcal{P}(\mathcal{S})
\end{equation}
is called a {\it random Gibbs measure} if for $P$-almost all $W$, $\mu(W)$ has a Markov property, standard for Gibbs measures, c.f. Definition 6.2.5 in \cite{[Bovier]}. The measurability in (\ref{Mes}) is the key point
since only in this case one can speak about {\it averages with respect to the disorder}, that is, about the expectations
$ \mathbb{E}_{P} \Phi \left( \mathbb{E}_{\mu(W)}F\right)$,
where $F:\mathcal{S}\rightarrow \mathbb{R}$ and $\Phi:\mathbb{R}\rightarrow \mathbb{R}$ are appropriate functions, see the discussion in Section 6.2 in \cite{[Bovier]}.
In general, for models with the interactions (\ref{O}), there might exist multiple Gibbs measures\footnote{The a.s. uniqueness of Gibbs measures of disordered spin systems is a highly nontrivial problem, see the discussion and the corresponding references in Section 6.3 in \cite{[Bovier]}.}. Hence, the map $W\mapsto \{\mu(W): \mu(W)$ is a Gibbs measure$\}$ can be set-valued  and the existence of its measurable selections (\ref{Mes}) is not obvious. To the best of our knowledge, in a systematic way  this aspect of the theory has never been discussed so far. Thus, one of  the aims of this work is to look at the problem of Gibbs fields with random interactions from the point of view of the set-valued analysis \cite{[Aubin]}. Another aim is to elaborate a method, which would allow to study also models with unbounded interactions -- the other irregular case that has not been studied yet.  In order to make the things as much  transparent as possible, we consider the simplest case where the graph is a lattice $\mathbb{Z}^d$ with the edge set ${\sf E} = \{(x,y) : |x-y|=1\}$, whereas the interaction potentials have the form
\begin{equation}
 \label{j3}
W_{x y} (u,v) =  J_{x y} u v, \quad J_{xy}, u ,v \in \mathbb{R},
\end{equation}
that is, all $S_x$ are the copies of $\mathbb{R}$.
In the physical terminology, this is a lattice spin model with unbounded spins  and a harmonic pair interaction. If all $J_{xy}$ are the same (or just uniformly bounded) and nonrandom, the existence and the properties of the corresponding Gibbs fields were studied since the 1970th, see \cite{[Ru],[LeP]} and the bibliographic notes in \cite{[P]}. However, the case of
$\sup_{\langle x,y \rangle \in {\sf E}}|J_{xy}| = + \infty$ has not been studied so far. To control the growth of $J= (J_{xy})_{\langle x,y\rangle \in {\sf E}}$ and $\sigma = (\sigma(x))_{x \in {\sf X}}$, we introduce two Banach spaces $\mathcal{J}_q \subset \mathbb{R}^{\sf E}$ and $\mathcal{S}_p\subset \mathbb{R}^{\mathbb{Z}^d}$. They are large enough so that every ball in
$\mathcal{J}_q$ contains $J $ with arbitrarily big $|J_{xy}|$. Then the interaction randomness  is realized as the triple $(\mathcal{J}_q, \mathcal{B}(\mathcal{J}_q), \nu)$, where $\nu$ is a general complete probability measure (need not be product, etc).
For every finite $\mathit{\Delta}\subset \mathbb{Z}^d$, by means of the potentials (\ref{j3}) we introduce the local conditional Gibbs measure $\pi_{\mathit{\Delta}} (\cdot| J, \xi)$, $J\in \mathcal{J}_q$ and $\xi\in \mathcal{S}_p$, which then allows us to
define the set of {\it tempered} Gibbs measures $\mathcal{G}_p(J)$ consisting of those $\mu \in \mathcal{P}(\mathbb{R}^{\mathbb{Z}^d})$ which solve the DLR equation and are such that $\mu(\mathcal{S}_p)=1$.
We prove that:
\vskip.1cm
\begin{tabular}{ll}
 (a) &for every $J\in \mathcal{J}_q$, the  set $\mathcal{G}_p(J)$ is non-void and weakly  compact,\\ & and that each $\mu \in \mathcal{G}_p (J)$ obeys an integrability estimate, the same \\ &for all such $\mu$  (Theorem \ref{1tm});\\[.2cm]
(b) &the map $\mathcal{J}_q\ni J \mapsto \mathcal{G}_p(J)$ is measurable, as a set-valued map, and\\ &hence there exist measurable selections $\mathcal{J}_q\ni J \mapsto \mu(J)\in \mathcal{G}_p(J)$\\ &(Theorem \ref{2tm}).
\end{tabular}
\vskip.1cm \noindent
The key element of the proof of (a) is an integrability estimate for the measures $\pi_{\mathit{\Delta}} (\cdot| J, \xi)$ that implies the existence of the accumulations points of the family $\{\pi_{\mathit{\Delta}} (\cdot| J, \xi)\}_{\mathit{\Delta}\subset \mathbb{Z}^d}$, which are elements of $\mathcal{G}_p(J)$. Then the corresponding estimate for $\mu \in \mathcal{G}_p(J)$, which holds uniformly for all such $\mu$ and all
$\| J \|_{q} \leq R$, $R>0$, are obtained therefrom.
This allows us to prove that the map
$\mathcal{J}_q\ni J \mapsto \mathcal{G}_p(J)$ is upper semi-continuous, which extends the result obtained (for bounded interactions) in item (d) of Theorem 4.23 in \cite{[Ge]}, page 72. By Theorem 8.1.4 of \cite{[Aubin]}, page 310, the mentioned
upper semi-continuity implies the measurability of
$\mathcal{J}_q\ni J \mapsto \mathcal{G}_p(J)$, which in turn yields the existence of measurable selections, see Theorem 8.1.3 in \cite{[Aubin]}.  In Corollary \ref{Coco}, we also establish the existence of the
averages $\mathbb{E}_{\nu} \Phi \left( \mathbb{E}_{\mu(J)}F\right)$ for appropriate functions $F$ and $\Phi$. Note  that the constants in (\ref{j14d}) are explicitly expressed in terms of the model parmeters.  

As is commonly accepted, see Chapter 7 in \cite{[Ge]}, the extreme elements of $\mathcal{G}_p(J)$ correspond to the thermodynamic phases of the physical system modeled by the family $\{\pi_{\mathit{\Delta}} (\cdot| J, \xi)\}_{\mathit{\Delta}\subset \mathbb{Z}^d}$. These elements are contained in the set of {\it limiting} Gibbs measures (Minlos states), see Corollary 7.30 on page 135 in \cite{[Ge]}, which are exactly the accumulation points of
$\{\pi_{\mathit{\Delta}} (\cdot| J, \xi)\}_{\mathit{\Delta}\subset \mathbb{Z}^d}$.  The physical meaning of such limiting Gibbs measures is that they approximate Gibbs measures of large finite systems, c.f. the corresponding discussion in \cite{[NS1]} and also in \cite{[Enter]}.  The random Gibbs measures obtained in (b) as measurable selections  need not be limiting Gibbs measures --  thus, the result of Theorem \ref{2tm} has rather theoretical value from the point of view of physics.
The characteristic feature of the spin models with random interactions is the so called  {\it chaotic dependence} of the measures $\pi_{\mathit{\Delta}} (\cdot| J, \xi)$ on $\mathit{\Delta}$, see \cite{[NS]} and the references cited therein. This means that the limits of the sequences   $\{\pi_{\mathit{\Delta}_n} (\cdot| J, \xi)\}_{n\in \mathbb{N}}$ need not be measurable (with respect to $J$) and hence cannot serve as limiting random Gibbs measures.
With the help of the Koml{\'o}s theorem \cite{[Komlos],[Balder]}, in  Theorem \ref{3tm} we obtain  that
\vskip.1cm
\begin{tabular}{ll}
 (c) &for every $\xi\in \mathcal{S}_p$, there exists a random Gibbs measure $\mu^\xi(J)$ and\\ &an exhausting sequence $\mathcal{D} = \{\mathit{\Delta}_n\}_{n\in \mathbb{N}}$ such that $\mu^\xi(J)$ is the $\nu$-a.s.\\ &weak limit of the sequence of `empirical distributions'
\end{tabular}
\begin{equation}
 \label{CC}
\frac{1}{N}\sum_{n=1}^N \pi_{\mathit{\Delta}_n}(\cdot|J,\xi), \qquad {N\in \mathbb{N}}.
\end{equation}
Under rather general assumptions,  each Gibbs measure has
the extreme decomposition, see Theorem 7.26 in \cite{[Ge]}, page 133. Thus, every measurable selection can be written in the form
\begin{equation}
 \label{geo}
\mu(J) = \int_{\mathcal{G}^{\rm ex}_p(J)} \mu \ \mathfrak{w}(J) (d\mu).
\end{equation}
Here $\mathcal{G}^{\rm ex}_p(J)$ is the extreme boundary of $\mathcal{G}_p(J)$ and  $\mathfrak{w}(J)$ is a {\it weight}, uniquely determined by $\mu(J)$. This decomposition holds for all $J\in \mathcal{J}_q$, but the weights $\mathfrak{w}(J)$ need not be $J$-measurable.  Suppose now that a representation holds which is similar to (\ref{geo}) with a measurable weight and the integral taken over the whole set $\mathcal{G}_p(J)$. Then it yields a random Gibbs measure and the corresponding weight
is called an
{\it Aizenman-Wehr metastate}, see e.g. page 103 in \cite{[Bovier]} and also Definition \ref{12df} below.
In Theorem \ref{4tm}, we show that
\vskip.1cm
\begin{tabular}{ll}
(d) &for every $\xi\in \mathcal{S}_p$, there exists an Aizenman-Wehr metastate $\mathfrak{m}^\xi (J)$\\ &and an  exhausting sequence $\mathcal{D} = \{\mathit{\Delta}_n\}_{n\in \mathbb{N}}$ such that $\mathfrak{m}^\xi (J)$ is the\\ &$\nu$-a.s. weak limit of the sequence of empirical metastates\\ &$\{N^{-1}\sum_{n=1}^N \delta_{\pi_{\mathit{\Delta}_n}(\cdot|J,\xi)}\}_{N\in \mathbb{N}}$.
\end{tabular}
\vskip.1cm \noindent
The {\it thermodynamic pressure}, or the free energy density, is an important characteristic which one obtains
in the thermodynamic limit, see. e.g. the discussion in \cite{[Bovier]}, pp. 24-28. For non-random (translation invariant) systems, the pressure exists and is independent of the way the limit has been taken, see Sections 2 and 3 in \cite{[LeP]} or Theorem 3.10 in \cite{[KP]}. In Theorem \ref{5tm}, we show that
under an additional condition on the measure $\nu$ the pressure can be obtained as the almost sure limit of the local pressures, `averaged' over $\{\mathit{\Delta}_n\}$ similarly as in (\ref{CC}). In Theorem \ref{6tm}, we assume that $\nu$ is a product measure with the zero first moment and prove that all the sequences of local pressures averaged over the disorder have one and the same thermodynamic limit -- the {\it quenched pressure}. In proving Theorems  \ref{5tm} and \ref{6tm}, we employ a version of the first GKS inequality, known for such models with $J_{xy} \geq 0$, which we obtain here
by extending the approach of \cite{[Cont],[Cont1]} to the case of unbounded interactions.

\section{Setup}
\subsection{General setting}

In constructing Gibbs random fields, we follow the standard scheme \cite{[Ge]}.
Our Gibbs fields will live on the set ${\sf X} = {\mathbb{Z}^d}$, $d\in \mathbb{N}$, equipped with the adjacency relation
$x \sim y$ defined by the condition $|x-y|=1$. By ${\sf E}$ we denote the set of edges of the corresponding
graph. We also use the shorthand
\[
\sum_{x} = \sum_{x\in \mathbb{Z}^d} , \ \ \ \ \sup_{x} = \sup_{x\in \mathbb{Z}^d}, \ \ \ \ \sum_{y\sim x} = \sum_{ y \in \mathbb{Z}^d: \ y\sim x}.
\]
The set $\mathbb{R}^{\mathbb{Z}^d}$ is equipped with the product topology, which turns it into a Polish space -- a separable completely metrizable topological space. Let $C_{\rm b} (\mathbb{R}^{\mathbb{Z}^d})$ be the Banach space of bounded continuous functions $f:\mathbb{R}^{\mathbb{Z}^d} \rightarrow \mathbb{R}$ with the norm
\[
 \|f\|_{\infty} = \sup_{\sigma \in \mathbb{R}^{\mathbb{Z}^d}} |f(\sigma)|.
\]
By means of $C_{\rm b} (\mathbb{R}^{\mathbb{Z}^d})$ we define the weak topology on the set of all probability measures $\mathcal{P}(\mathbb{R}^{\mathbb{Z}^d})$, which turns it into a Polish space, see e.g. page 39 in \cite{[Part]}.

For any $\mathit{\Delta} \subset \mathbb{Z}^d$, we let $\mathit{\Delta}^c \ \stackrel{\rm def}{=} \ \mathbb{Z}^d \setminus \mathit{\Delta}$; by writing
$\mathit{\Delta} \Subset {\sf X}$ we mean that   $0<
|\mathit{\Delta}|<\infty $. A sequence $\mathcal{D} = \{\mathit{\Delta}_n \}_{n\in \mathbb{N}}$, such that $\mathit{\Delta}_n \Subset \mathbb{Z}^d$ for all $n\in \mathbb{N}$, is said to be {\it cofinal} if it is: (a)
ordered by inclusion; (b) exhausting, i.e. such that each $x\in \mathbb{Z}^d$ belongs to a certain $\mathit{\Delta}_n$.
For $\mathit{\Delta}\subset \mathbb{Z}^d$, by $\mathcal{B}_{\mathit{\Delta}}$ we denote
the $\sigma$-sub-field of $\mathcal{B}(\mathbb{R}^{\mathbb{Z}^d})$ generated by $(\sigma(x))_{x\in \mathit{\Delta}}$. For
$\mathit{\Delta}\Subset \mathbb{Z}^d$, a probability kernel $\pi_{\mathit{\Delta}} (\cdot| \cdot)$ is a function on
$(\mathcal{B}(\mathbb{R}^{\mathbb{Z}^d}), \mathbb{R}^{\mathbb{Z}^d})$ such that for any $\xi \in \mathbb{R}^{\mathbb{Z}^d}$,
$\pi_{\mathit{\Delta}}(\cdot | \xi)$ is in $\mathcal{P}(\mathbb{R}^{\mathbb{Z}^d})$, and for any $A \in \mathcal{B}(\mathbb{R}^{\mathbb{Z}^d})$,
$\pi_{\mathit{\Delta}}(A | \cdot )$ is $\mathcal{B}_{\mathit{\Delta}^c}$-measurable. Such a kernel is said to
be proper if $\pi_{\mathit{\Delta}} (A| \cdot) = \mathbb{I}_A(\cdot)$ for any $A \in
\mathcal{B}_{\mathit{\Delta}^c}$. Here $\mathbb{I}_A(\xi) = 1$ if $\xi \in A$, and $\mathbb{I}_A(\xi) = 0$
otherwise. Given a family $\{\pi_{\mathit{\Delta}}\}_{\mathit{\Delta}\Subset \mathbb{Z}^d}$, suppose that there exists
$\mu \in \mathcal{P}(\mathbb{R}^{\mathbb{Z}^d})$ such that
\begin{equation}
 \label{i1}
\mu(A |\mathcal{B}_{\mathit{\Delta}^c}) = \pi_{\mathit{\Delta}} (A| \cdot),
\end{equation}
which holds $\mu$-almost surely for all $A\in \mathcal{B}(\mathbb{R}^{\mathbb{Z}^d})$ and $\mathit{\Delta} \Subset \mathbb{Z}^d$. Then this measure
$\mu$ is said to be specified by the family $\{\pi_{\mathit{\Delta}}\}_{\mathit{\Delta}\Subset \mathbb{Z}^d}$. In this
case, all the kernels $\pi_{\mathit{\Delta}}$ are $\mu$-almost surely proper, and their family is $\mu$-almost surely consistent. The latter means that for $\mu$-almost all $\xi$ and all $A\in \mathcal{B}(\mathbb{R}^{\mathbb{Z}^d})$,
\begin{equation}
 \label{i2}
\int_{\mathbb{R}^{\mathbb{Z}^d}} \pi_{\mathit{\Lambda}} (A| \eta) \pi_{\mathit{\Delta}} (d \eta | \xi) =
\pi_{\mathit{\Delta}} (A| \xi),
\end{equation}
which holds for any pair of subsets such that $\mathit{\Lambda} \subset \mathit{\Delta}$. It should be pointed out that (\ref{i1}) is equivalent to
\begin{equation}
  \label{i0}
  \int_{\mathbb{R}^{\mathbb{Z}^d}} \pi_{\mathit{\Delta}} (A|\xi) \mu(d \xi) = \mu(A),
\end{equation}
which holds for all $A\in \mathcal{B}(\mathbb{R}^{\mathbb{Z}^d})$ and $\mathit{\Delta}\Subset {\mathbb{Z}^d}$. The condition (\ref{i0})  is called the {\it Dobrushin-Lanford-Ruelle (DLR)} equation. It is equivalent to
\begin{equation}
  \label{jj0}
  \int_{\mathbb{R}^{\mathbb{Z}^d}} \pi_{\mathit{\Delta}} (f|\xi) \mu(d \xi) = \mu(f),
\end{equation}
satisfied for all $f \in C_{\rm b}(\mathbb{R}^{\mathbb{Z}^d})$ and all $\mathit{\Delta}\Subset {\mathbb{Z}^d}$.
Here we use the notation
\begin{equation}
 \label{jj1}
\mu (f) =   \int_{\mathbb{R}^{\mathbb{Z}^d}} f(\sigma) \mu(d \sigma).
\end{equation}
\subsection{The Gibbs fields}

The Gibbs fields we are going to construct are specified by the kernels obtained as perturbations of
the products of single--spin measures by the factors (\ref{O}) with the functions $W_{x y}$ as in (\ref{j3}).
For $\mathit{\Delta} \Subset \mathbb{Z}^d$ and $\xi \in \mathbb{R}^{\mathbb{Z}^d}$, we set
\begin{eqnarray}
 \label{j4}
- H_{\mathit{\Delta} } ( \sigma_{\mathit{\Delta}} |J,\xi) & = & \sum_{\langle x, y \rangle \in {\sf E}_{\mathit{\Delta}} } J_{xy} \sigma (x) \sigma (y) \\[.2cm] & + & \sum_{x \in \mathit{\Delta}, \ y\in \mathit{\Delta}^c, \ x \sim y} J_{xy} \sigma (x)  \xi (y), \nonumber
\end{eqnarray} where ${\sf E}_{\mathit{\Delta}}$ consists  of the edges with both endpoints in $\mathit{\Delta}$. In the mentioned terminology, $H_{\mathit{\Delta} } (\sigma_{\mathit{\Delta}} |J,\xi)$ is
the energy of the interaction of the spins located in $\mathit{\Delta}$ with each other and with the fixed spins outside $\mathit{\Delta}$. For a family $\chi = (\chi_x)_{x\in {\mathbb{Z}^d}}$, $\chi_x \in \mathcal{P}(\mathbb{R})$, we put
\begin{equation}
  \label{i6}
  \chi_{\mathit{\Delta}} (d \sigma_{\mathit{\Delta}} ) = \prod_{x\in \mathit{\Delta}} \chi_x (d \sigma (x)), \quad \sigma_{\mathit{\Delta}} = (\sigma(x))_{x\in \mathit{\Delta}},
\end{equation}
which is an element of $\mathcal{P}(\mathbb{R}^{|\mathit{\Delta}|})$.
Thereafter, for $A \in \mathcal{B}(\mathbb{R}^{\mathbb{Z}^d})$, we define
\begin{eqnarray}
 \label{j5}
& & \pi_{\mathit{\Delta}} ( A|J,\xi)\\[.2cm] & & \qquad  =  \frac{1}{Z_{\mathit{\Delta}} (J,\xi)} \int_{\mathbb{R}^{|\mathit{\Delta}|}}
\mathbb{I}_A ( \sigma_{\mathit{\Delta}} \times \xi_{\mathit{\Delta}^c} ) \exp[ - H_{\mathit{\Delta}} ( \sigma_{\mathit{\Delta}} |J,\xi)] \chi_{\mathit{\Delta}} (d \sigma_{\mathit{\Delta}}). \nonumber
\end{eqnarray}
Here $Z_{\mathit{\Delta}} (J,\xi)$ is a normalizing factor, that is,
\begin{equation}
 \label{j5b}
Z_{\mathit{\Delta}} (J,\xi) = \int_{\mathbb{R}^{|\mathit{\Delta}|}}
 \exp[ - H_{\mathit{\Delta}} ( \sigma_{\mathit{\Delta}} |J,\xi)] \chi_{\mathit{\Delta}} (d \sigma_{\mathit{\Delta}}),
\end{equation} and the juxtaposition stands for the element of
$\mathbb{R}^{\mathbb{Z}^d}$ such that
\[
(\sigma_{\mathit{\Delta}} \times \xi_{\mathit{\Delta}^c} ) (x) = \sigma (x), \  {\rm for} \ x\in \mathit{\Delta}; \quad \ (\sigma_{\mathit{\Delta}} \times \xi_{\mathit{\Delta}^c} ) (x) = \xi (x), \  {\rm for} \ x\in \mathit{\Delta}^c.
\]
The family $\{\pi_{\mathit{\Delta}}\}_{\mathit{\Delta}\Subset \mathbb{Z}^d}$ is clearly consistent. It is the local Gibbs specification for our model.

As is typical for Gibbs measures of models with unbounded spins, the description of the properties possessed by all such measures is rather unrealistic. Usually, the study is restricted to those measures which have a prescribed support property. Such measures are called {\it tempered}. To define the mentioned property we use a weight function $w:\mathbb{Z}^d\rightarrow (0,1]$, which by definition has the following properties:
\begin{eqnarray}
  \label{j8}
& (a) & \qquad \qquad \ \  |w| \ \stackrel{\rm def}{=} \ \sum_{x} w(x) < \infty,\\[.2cm]
 \label{j9}
& (b) & \qquad \exists w_0 >0 \quad
 w(x) \leq w_0 w(y), \quad {\rm for} \  {\rm all} \ \  x\sim y.
\end{eqnarray}
Note that $w_0\geq 1$, otherwise one would get $w(x) \equiv 0$.
A typical example can be
\begin{equation}
\label{exa}
w (x) = \exp(-\alpha |x|), \quad \alpha >0.
\end{equation}
For $w$ obeying (\ref{j8}) and (\ref{j9}) and for a $p\geq1$, we set
\begin{equation}
  \label{j10}
 \|\sigma \|_p = \left(\sum_{x} |\sigma (x) |^p w(x) \right)^{1/p},
\end{equation}
and
\begin{equation}
  \label{j11}
 \mathcal{S}_p= L^p (\mathbb{Z}^d, w) = \{ \sigma \in \mathbb{R}^{\mathbb{Z}^d} \ : \ \|\sigma \|_p < \infty\}.
\end{equation}
This will be the space of tempered spin configurations. Next, for $q\geq 1$, we introduce the space of tempered interaction intensities
\begin{eqnarray}
  \label{j12}
& & \|J \|_{q}  =  \left(\sum_{\langle x,y \rangle \in {\sf E}} |J_{xy}|^q [w(x) + w(y)] \right)^{1/q}, \\[.2cm]
 & & \mathcal{J}_q= L^q ({\sf E}, w)  =  \{ J \in \mathbb{R}^{\sf E} \ : \ \|J \|_q < \infty\} . \nonumber
 \end{eqnarray}
Clearly,  $L^p (\mathbb{Z}^d, w)$ and $L^q ({\sf E}, w)$ are measurable subsets of the Polish spaces $\mathbb{R}^{\mathbb{Z}^d}$ and $\mathbb{R}^{\sf E}$, respectively. We equip these sets with the corresponding norm topologies, which turns them into separable Banach spaces. It can easily be shown (see also the Kuratowski theorem, page 15 in \cite {[Part]}), that
\begin{equation} \label{j120}
 \mathcal{B}(\mathcal{S}_p) = \{ \mathcal{S}_p\cap A : A \in \mathcal{B} (\mathbb{R}^{\mathbb{Z}^d})\}.
\end{equation}
Thus, one can consider the set
\begin{equation}
  \label{j12a}
\mathcal{P}_{\rm temp}  =  \{ \mu \in \mathcal{P} (\mathbb{R}^{\mathbb{Z}^d}) : \mu\left( \mathcal{S}_p \right) =1\}.
\end{equation}
The elements of $\mathcal{P}_{\rm temp}$ are called tempered measures.
Now we impose conditions on the family of single-spin measures $\chi = (\chi_x)_{x\in \mathbb{Z}^d}$.
For $\lambda >0$ and $q>1$, we set
\begin{eqnarray}
  \label{j13}
& & \sup_{x}\int_{\mathbb{R}} \exp\left( \lambda |u|^{2q/(q-1)}\right)\chi_x (d u) = C_+ (\lambda),\\[.2cm]
& & \inf_{x}\int_{\mathbb{R}} \exp\left(- \lambda |u|^{2q/(q-1)}\right)\chi_x (d u) = C_- (\lambda).\nonumber
\end{eqnarray}
And then
\begin{equation}
  \label{j14}
\mathcal{K}_q \ \stackrel{\rm def}{=} \ \{ \chi = (\chi_x)_{x\in \mathbb{Z}^d} : \forall \lambda >0 \ \ C_{+}( \lambda) < \infty , \  C_{-}(\lambda) >0\}.
\end{equation}
As an example of $\chi \in \mathcal{K}_q $ one can take the  copies of the measure
$\chi_0 (du)\sim \exp(- V(u)) du$, where $V$ is an even semi-bounded polynomial of ${\rm deg} V > 2q/(q-1)$, c.f. \cite{[LeP],[P]}. This corresponds to the physical model called an {\it anharmonic crystal} where the spins are the displacements of the oscillators from their equilibrium positions.

In the sequel, we shall always choose $J$ in $\mathcal{J}_q$ and  $\chi$ in $\mathcal{K}_q$ with one and the same $q>1$.
We also assume that this $q$ and $p$ in (\ref{j11}) and (\ref{j12a}) satisfy
\begin{equation}
  \label{j14a}
  p = \frac{2q}{q-1},
\end{equation}
i.e. $p>2$.
As the main our concern is the dependence on $J$, the dependence on $\chi$ will always be suppressed from the notations.
\begin{definition}
  \label{1df}
Given $J = (J_{xy})_{\langle x,y\rangle \in {\sf E}}\in \mathcal{J}_q$ and $p$ as in (\ref{j14a}), by  $\mathcal{G}_p(J)$ we denote the set of all
$\mu\in \mathcal{P}_{\rm temp}$ which solve the DLR equation (\ref{i0}) with the kernels defined in (\ref{j4}) and (\ref{j5}). The elements of  $\mathcal{G}_p(J)$ are called (tempered) Gibbs measures.
\end{definition}
We recall that a probability space $(\Omega, \mathcal{O}, P)$ is said to be {\it complete} if for every $A$ such that $P(A) = 0$, each subset of $A$ is in $\mathcal{O}$. We also recall that $\mathcal{J}_q$ is a separable Banach space.
\begin{definition}
  \label{10df}
By the lattice model with unbounded spins and unbounded random interactions we mean the pair
\[(\mathcal{J}_q, \mathcal{B}(\mathcal{J}_q), \nu) \quad {\rm and} \quad \{\pi_{\mathit{\Delta}}(\cdot |J,\xi): \mathit{\Delta} \Subset \mathbb{Z}^d, \ J\in \mathcal{J}_q, \ \xi \in \mathcal{S}_p \},\] where the probability space is complete and the kernels $\pi_{\mathit{\Delta}}$ are defined in (\ref{j4}) and (\ref{j5}).
\end{definition}
\begin{definition}
 \label{11df}
A $\mathcal{B}(\mathcal{J}_q)/\mathcal{B}(\mathcal{P}(\mathbb{R}^{\mathbb{Z}^d}))$-measurable map $\mathcal{J}_q \ni J \mapsto \mu (J) \in \mathcal{P}(\mathbb{R}^{\mathbb{Z}^d})$ is said to be a random Gibbs measure if $\mu(J) \in \mathcal{G}_p (J)$ for $\nu$-almost all $J\in \mathcal{J}_q$.
\end{definition}
Note that when we speak about a Gibbs measure $\mu$ we mean merely an element of a given $\mathcal{G}_p(J)$. However,
a {\it random Gibbs measure} $\mu(J)$ will stand for a measure-valued function of $J\in \mathcal{J}_q$.

Let
$\mathfrak{P}$ denote the space of all probability measures on
$(\mathcal{P}(\mathbb{R}^{\mathbb{Z}^d}),\mathcal{B}(\mathcal{P}(\mathbb{R}^{\mathbb{Z}^d})))$.
We equip it with the weak topology and thereby with the Borel $\sigma$-field $\mathfrak{B}$.
For every $f\in C_{\rm b} (\mathbb{R}^{\mathbb{Z}^d})$, the evaluation map $\mathcal{P}(\mathbb{R}^{\mathbb{Z}^d}) \ni \mu \mapsto \mu(f) $
is continuous and bounded.
\begin{definition}
 \label{12df}
A $\mathcal{B}(\mathcal{J}_q)/\mathfrak{B}$-measurable map $\mathcal{J}_q \ni J \mapsto \mathfrak{m} (J) \in \mathfrak{P}$ is said to be an Aizenman-Wehr metastate if
\vskip.1cm
\begin{tabular}{ll}
(a) &$\mathfrak{m} (J) \left(\mathcal{G}_p (J) \right)=1$ for $\nu$-almost all $J\in \mathcal{J}_q$;\\[.2cm]
(b) &the map
\end{tabular}
\begin{equation} \label{dfg}
 \mathcal{J}_q \ni J \mapsto \int_{\mathcal{P}(\mathbb{R}^{\mathbb{Z}^d})} \mu \ \mathfrak{m} (J) (d \mu) \in  \mathcal{P}(\mathbb{R}^{\mathbb{Z}^d})
\end{equation}
\hskip1.8cm is a random Gibbs measure.
\end{definition}
Note that the integral in (\ref{dfg}) is understood in terms of the pairing with
$f\in C_{\rm b} (\mathbb{R}^{\mathbb{Z}^d})$.

\section{The results}

\subsection{Theorems }
For $R>0$, we set $B_q(R) = \{J \in \mathcal{J}_q : \|J\|_q \leq R\}$. From (\ref{j12}) it follows that
\begin{equation}
 \label{lack}
\sup_{J\in B_q(R)} \sup_{\langle x,y \rangle \in {\sf E}} |J_{xy}| = +\infty,
\end{equation}
for any $R>0$. Recall that the set of tempered measures $\mathcal{P}_{\rm temp}$ was defined in  (\ref{j12a}). In the sequel, when we discuss topological properties of $\mathcal{G}_p (J)$ we always mean the topology induced by the weak topology of the space $\mathcal{P}(\mathbb{R}^{\mathbb{Z}^d})$ (defined by means of $C_{\rm b}(\mathbb{R}^d)$).
\begin{theorem}
  \label{1tm}
For every $J\in \mathcal{J}_q$, $q>1$, the set $\mathcal{G}_p(J)$ ($p$ as in (\ref{j14a}))
is non-void and compact. For any $\lambda >0$, there exist positive constants $\mathit{\Upsilon}_i(\lambda)$, $i=1,2$, such that for every $\mu\in \mathcal{G}_p(J)$, the following estimate holds
\begin{equation}
  \label{J14aa}
\int_{\mathbb{R}^{\mathbb{Z}^d}} \exp\left(\lambda \|\sigma\|^p_p  \right)\mu(d\sigma) \leq \exp\left( \mathit{\Upsilon}_1 (\lambda) + \mathit{\Upsilon}_2 (\lambda)\|J\|_q^q\right).
\end{equation}
\end{theorem}
By Jensen's inequality, one readily gets from (\ref{J14aa}) the  next
\begin{corollary}
 \label{Coco}
There exist
positive constants $A$ and $B$ such that for any random Gibbs measure
$\mu(J)$, the following estimate
\begin{equation}
  \label{j14d}
\int_{\mathcal{J}_q} \Phi \left(\int_{\mathbb{R}^{\mathbb{Z}^d}}  \|\sigma\|^p_p\mu(J)(d\sigma) \right) \nu (d J) \leq \int_{\mathcal{J}_q}
\Phi\left( A + B\|J\|_q^q \right) \nu (d J)
\end{equation}
holds for any increasing function $\Phi:\mathbb{R}_+\rightarrow \mathbb{R}_+$.
\end{corollary}
Random Gibbs measures can be obtained as measurable selections.
\begin{definition}
 \label{Medf}
A measurable map $\mathcal{J}_q \ni J \mapsto \mu(J) \in \mathcal{P}(\mathbb{R}^{\mathbb{Z}^d})$ such that
\[
 \forall J \in \mathcal{J}_q : \quad \mu(J) \in \mathcal{G}_p(J)
\]
is called a measurable selection of the set-valued map $\mathcal{J}_q \ni J \mapsto \mathcal{G}_p(J)\subset \mathcal{P}(\mathbb{R}^{\mathbb{Z}^d})$.
\end{definition}
\begin{theorem}
  \label{2tm}
Let $p$ and $q$ be as in Theorem \ref{1tm} and the probability space $(\mathcal{J}_q, \mathcal{B}(\mathcal{J}_q), \nu)$ be as in  Definition \ref{10df}. Then the map $\mathcal{J}_q \ni J \mapsto \mathcal{G}_p(J)$ has measurable selections.  \end{theorem}
It turns out that measurable selections constitute quite a big subset of the set of Gibbs measures, c.f. item (vi) of Theorem 8.1.4 in \cite{[Aubin]}, page 310.
\begin{remark}
 \label{aubinrk}
There exists an at most countable family $\{\mu_n\}_{n\in \mathbb{N}}$ of measurable selections mentioned in Theorem \ref{2tm}  such that, for every $J\in \mathcal{J}_q$, the set $\{\mu_n(J)\}_{n\in \mathbb{N}}\subset \mathcal{G}_p(J)$ is dense in $\mathcal{G}_p(J)$. Thus, $\mathcal{G}_p(J)$  is a singleton for $\nu$-almost all $J$ if there is only one measurable selection.
\end{remark}
As was already mentioned in Introduction, only limiting Gibbs measures can serve as the approximations of the Gibbs measures of large finite systems, see \cite{[NS1]}. In the next theorem, we obtain random Gibbs measures  as weak limits of the averaged kernels $\pi_{\mathcal{D},N}$.
For a cofinal sequence $\mathcal{D}=\{\mathit{\Delta}_n\}_{n\in \mathbb{N}}$ and $N\in \mathbb{N}$, we set
\begin{equation}
 \label{Co1}
\pi_{\mathcal{D},N} (\cdot |J, \xi) = \frac{1}{N} \sum_{n=1}^N \pi_{\mathit{\Delta}_n} (\cdot |J, \xi).
\end{equation}
\begin{theorem}
 \label{3tm}
For every $\xi\in \mathcal{S}_p$, there exists a random Gibbs measure $\mu^\xi$ and a cofinal sequence $\mathcal{D}$ such that, in the topology of $\mathcal{P}(\mathbb{R}^{\mathbb{Z}^d})$,  one has $\mu^\xi (J) = \lim_{N\rightarrow + \infty} \pi_{\mathcal{D},N} (\cdot|J, \xi)$ for $\nu$-almost all $J\in \mathcal{J}_q$.
\end{theorem}
The fact that for approximating finite volume Gibbs measures we can use the sequences of averaged kernels rather than the sequences of kernels themselves can be explained by the chaotic dependence of the kernels $\pi_{\mathit{\Delta}} (\cdot |J, \xi)$ on $\mathit{\Delta}$, which is smoothed up in (\ref{Co1}).

For $\mathit{\Delta} \Subset \mathbb{Z}^d$, we let
$\mathfrak{d}^\xi_{\mathit{\Delta}} (J)$ denote the $\delta$-measure centered at $\pi_{\mathit{\Delta}} (\cdot|J, \xi)$, that is $\mathfrak{d}^\xi_{\mathit{\Delta}} (J)(A) = \mathbb{I}_A \left(\pi_{\mathit{\Delta}} (\cdot|J, \xi) \right)$ for all $A\in \mathfrak{B}$. Then for a cofinal sequence $\mathcal{D}$ and $N\in \mathbb{N}$, we set, c.f. (\ref{Co1}),
\begin{equation}
 \label{Co2}
\mathfrak{d}^\xi_{\mathcal{D},N} (J) = \frac{1}{N} \sum_{n=1}^N \mathfrak{d}^\xi_{\mathit{\Delta}_n} (J),
\end{equation}
which is the {\it Newman-Stein empirical metastate}, see eq. (B19) on page 77 in \cite{[Newman]} or eq. (A18) on page 281 in \cite{[NS]}. Recall that the Aizenman-Wehr metastates were introduced in Definition \ref{12df}.
\begin{theorem}
 \label{4tm}
For every $\xi\in \mathcal{S}_p$, there exists an Aizenman-Wehr metastate $\mathfrak{m}^\xi$ and a cofinal sequence $\mathcal{D}$ such that, in the topology of $\mathfrak{P}$, $\mathfrak{m}^\xi (J) = \lim_{N\rightarrow +\infty} \mathfrak{d}^\xi_{\mathcal{D},N} (J)$ for $\nu$-almost all $J\in \mathcal{J}_q$.
\end{theorem}
For $\mathit{\Delta}\Subset \mathbb{Z}^d$, $J\in \mathcal{J}_q$, and $\xi \in \mathcal{S}_p$, the (local) {\it pressure} in $\mathit{\Delta}$ is
\begin{equation}
 \label{pres}
p_{\mathit{\Delta}} (J, \xi) = \frac{1}{|\mathit{\Delta}|} \log Z_{\mathit{\Delta}} (J, \xi),
\end{equation}
where $Z_{\mathit{\Delta}} (J, \xi)$ is the same as in (\ref{j5b}).  Like in (\ref{Co1}),  for a cofinal sequence $\mathcal{D}= \{\mathit{\Delta}_n\}_{n\in \mathbb{N}}$ and $N\in \mathbb{N}$, we consider
\begin{equation}
 \label{pres1}
p_{\mathcal{D},N} (J, \xi) = \frac{1}{N} \sum_{n=1}^N p_{\mathit{\Delta}_n} (J, \xi).
\end{equation}
Let now $\mu$ be a
random Gibbs measure, see Definition \ref{11df}.
Then
\begin{eqnarray}
 \label{press}
\bar{p}^\mu_{\mathit{\Delta}} (J) & = & \int_{\mathbb{R}^{\mathbb{Z}^d}}p_{\mathit{\Delta}} (J, \xi) \mu(J)(d \xi), \\[.2cm] \bar{p}^\mu_{\mathcal{D},N} (J) & = & \int_{\mathbb{R}^{\mathbb{Z}^d}}p_{\mathcal{D},N} (J, \xi) \mu(J)(d \xi),\nonumber
\end{eqnarray}
are measurable functions of $J\in \mathcal{J}_q$, and
\begin{equation}
 \label{pres2}
\vartheta(d \sigma, d J) = \mu(J)(d \sigma) \nu(dJ)
\end{equation}
is a probability measure on the product space $\mathbb{R}^{\mathbb{Z}^d} \times \mathcal{J}_q$.
\begin{theorem}
 \label{5tm}
Suppose that $\nu$ has the property
\begin{equation}
 \label{pres3}
\sup_{\langle x,y\rangle \in {\sf E}}\int_{\mathcal{J}_q} |J_{xy}|^q \nu(d J) = a_\nu < +\infty.
\end{equation}
Then, for any random Gibbs measure $\mu$,
there exists a cofinal sequence $\mathcal{D}$ such that the sequence $\{\bar{p}^\mu_{\mathcal{D},N} (J)\}_{N\in \mathbb{N}}$ converges, for
$\nu$-almost all $J\in \mathcal{J}_q$, to a certain
$p^\mu\in L^1 (\mathcal{J}_q,\nu)$. Furthermore, for $\nu$ obeying (\ref{pres3}), let
$\vartheta$ be as in (\ref{pres2}). Then there exists a cofinal sequence $\mathcal{D}$ such that the sequence $\{p_{\mathcal{D},N} (J, \xi)\}_{N\in \mathbb{N}}$ converges, for
$\vartheta$-almost all $(\xi, J)\in \mathbb{R}^{\mathbb{Z}^d} \times \mathcal{J}_q$, to a certain
$p\in L^1 (\mathbb{R}^{\mathbb{Z}^d} \times \mathcal{J}_q,\vartheta)$.
\end{theorem}
Under one more condition on the measure $\nu$ we can strengthen the above result as follows.
A cofinal sequence $\mathcal{D} = \{\mathit{\Delta}_n\}_{n\in \mathbb{N}}$ is called a van Hove sequence if
\[
 \inf_{n\in \mathbb{N}} \frac{|\partial \mathit{\Delta}_n|}{ | \mathit{\Delta}_n|} = \lim_{n\rightarrow +\infty} \frac{|\partial \mathit{\Delta}_n|}{ | \mathit{\Delta}_n|} = 0,
\]
see e.g. page 193 in \cite{[A]}. Here $\partial \mathit{\Delta} = \{ y \in \mathit{\Delta}^c \ : \ \exists x\in \mathit{\Delta} \ \ x \sim y\}$.
\begin{theorem}
\label{6tm}
In addition to (\ref{pres3}), assume that  $\nu$ is a product measure such that
\begin{equation}
 \label{pres4}
\int_{\mathcal{J}_q} J_{xy} \nu(d J) = 0,
\end{equation}
for all $\langle x,y\rangle \in {\sf E}$. Then, for any cofinal sequence $\mathcal{D}= \{\mathit{\Delta}_n\}_{n\in \mathbb{N}}$, there exists the quenched pressure
\begin{equation}
 \label{pres5}
p^{\rm quen} =  \lim_{n\rightarrow +\infty} \int_{\mathcal{J}_q}p_{\mathit{\Delta}_n} (J,0)\nu (dJ) = \sup_{\mathit{\Delta} \Subset \mathbb{Z}^d} \int_{\mathcal{J}_q} p_{\mathit{\Delta}} (J,0)\nu (dJ),
\end{equation}
which thereby is independent of $\mathcal{D}$.
Furthermore, for any random Gibbs measure $\mu$ and any van Hove sequence $\mathcal{D}= \{\mathit{\Delta}_n\}_{n\in \mathbb{N}}$, we have that
\begin{equation}
 \label{pres6}
p^{\rm quen} =  \lim_{n\rightarrow +\infty} \int_{\mathcal{J}_q}\bar{p}^\mu_{\mathit{\Delta}_n} (J)\nu (dJ).
\end{equation}
\end{theorem}

\subsection{Comments}

All the results presented above can readily be extended to any bounded degree graph, and, after some modifications, also to unbounded degree graphs of a certain kind \cite{[KKP]}. They can also be extended to more general pair interaction potentials $W_{xy}$, c.f. (\ref{j3}). The only conditions would be the continuity as in Lemma \ref{New2lm} and that the interaction energy (\ref{j4}) obeys (\ref{een}) with appropriate $J= (J_{xy})\in \mathcal{J}_q$.
If every single-spin measure $\chi_x$ is supported on a bounded $[a,b]$, then all the results formulated
above hold true with any $q$ and $p=2q/ (q-1)$, including $q=1$ and $p=\infty$. In this case, we deal with regular random interactions\footnote{See Definition 6.2.1 in \cite{[Bovier]}}. An important particular model of this kind is
the Edwards-Anderson spin glass, see Section 2 in \cite{[NS]}.
In  this  model,  the  spins take values $\pm 1$ with equal probabilities and the interaction intensities $J_{xy}$ are symmetric, typically Gaussian, i.i.d..
Note that such a  model meets the conditions of Theorem \ref{6tm}.

More specific remarks to the above results
are as follows:
\begin{itemize}
\item {\bf Theorem \ref{1tm}.} The main point of this theorem is the lack of the uniform boundedness of the intensities $J_{xy}$, c.f. (\ref{lack}). Clearly, the growth of $J_{xy}$ should be controlled in one or another way. We do this by imposing the tempredness condition $\|J\|_q< \infty$, which appears in the right-hand side of (\ref{J14aa}) and in similar estimates.
The same results can be obtained for the Euclidean Gibbs measures which describe equilibrium thermodynamic states of lattice systems of interacting quantum anharmonic oscillators with random interactions. In this case, our Theorem \ref{1tm} would be an extension of Theorems 3.1 and 3.2 of \cite{[KP]} and of Theorems 3.3.1 and 3.3.6, pp. 214 - 216 in \cite{[A]}. For the Euclidean Gibbs measures, the single-spin spaces $S_x$ are the copies of the space of periodic continuous functions $\sigma_x : [0,\beta] \rightarrow \mathbb{R}$, where $\beta>0$ is the inverse temperature. In view of this, one needs to apply more sophisticated methods of the path integral approach \cite{[A]}.
\item {\bf Theorem \ref{2tm}.} If $J$ is random and fixed, the set $\mathcal{G}_p(J)$ describes the equilibrium thermodynamic
states of the spin system with {\it quenched} disorder. In order to average over the disorder, one has to have the measurability as in Definition \ref{11df}.
In Theorem \ref{1tm}, we prove that $\mathcal{G}_p(J)$ is non-void by showing that the family $\{\pi_{\mathit{\Delta}}(\cdot|J, \xi)\}_{\mathit{\Delta} \Subset \mathbb{Z}^d}$ possesses accumulation points, which are tempered Gibbs measures. Each such a measure is therefore obtained as the limit of $\{\pi_{\mathit{\Delta}_n}(\cdot|J, \xi)\}_{n \in  \mathbb{N}}$ for the corresponding sequence $\{\mathit{\Delta}_n\}_{n\in \mathbb{N}}$ which, however, can  depend on $J$ in an uncontrollable way (the so called {\it chaotic size dependence\footnote{See the discussion in  \cite{[NS]} and in \cite{[Newman]}, pp. 55, 56, 64.}}).
In view of this fact, it is unclear  whether these limiting points provide the measurability of $J\mapsto \mu(J)$.
In  Theorem \ref{2tm},  this measurability is obtained by means of the general methods of the set-valued analysis.
To the best of our knowledge, this is the first instance of the use of such methods in the theory of lattice models with random interactions.
\item {\bf Theorems \ref{3tm} and \ref{4tm}.} These theorems give a constructive procedure of obtaining random Gibbs measures as the infinite volume limits. Even for $p=\infty$ and $q=1$, i.e. in the regular case of bounded interactions,  Theorems \ref{3tm} and \ref{4tm} are the corresponding extensions of Theorems 6.2.6 and 6.2.8 in \cite{[Bovier]}, pp. 101--104.
The novelty of these our theorems is that
the chaotic size dependence is harnessed with the help of the Koml\'os theorem \cite{[Komlos]} -- a renowned tool in the probability theory. This provides a new look at the approach put forward by C. M. Newman and D. L. Stein, see \cite{[Newman],[NS],[NS1],[Bovier]} and the references therein.
\item {\bf Theorems \ref{5tm} and \ref{6tm}.} For the translation invariant lattice systems with nonrandom interactions, the thermodynamic pressure exists and is unique even if the Gibbs measures are multiple, see Theorem 3.10 and Corollary 3.11 in \cite{[KP]}, and/or Theorems 5.1.2 and 5.1.3 in \cite{[A]}. It is thus an important thermodynamic function by means of which one can establish e.g. the absence/existence of phase transitions, see \cite{[Ali]} and/or Chapter 6 in \cite{[A]}.
For the disordered systems, the pressure in $\mathit{\Delta}\Subset \mathbb{Z}^d$ clearly manifests the  chaotic size dependence.
For the model considered here, we propose to eliminate this effect by passing to the averages (\ref{pres1}), as
we did in Theorems \ref{3tm} and \ref{4tm}. The existence of the limiting quenched pressure obtained in (\ref{pres5}) is a generalization to unbounded spins of the relevant result of \cite{[Cont],[Cont1]}. The important point in Theorem \ref{6tm} is that the pressure averaged over the disorder is the same in all states, which resembles the corresponding fact known for nonrandom interactions, see Theorem 3.10 in \cite{[KP]} and Theorem 5.1.3 in \cite{[A]}, page 268. One observes that this result holds true also for the Edwards-Anderson spin glass.
For the systems of quantum anharmonic oscillators  with the corresponding random interactions, the analogous statements can readily be proven by means of a combination of the methods of \cite{[A],[KP]} and those of the present work.
This would be the extension of the
results of \cite{[ContL]}.
\end{itemize}

\section{The proof of the theorems}

In the next subsection, we formulate the lemmas which are then used to prove
Theorems \ref{1tm} and \ref{2tm}. They describe the basic (regularity) properties of the family
of local Gibbs measures $\{\pi_{\mathit{\Delta}}\}_{\mathit{\Delta}\Subset \mathbb{Z}^d}$.
The proof of these lemmas will be done in the next section.

\subsection{The basic lemmas}

In the lemmas formulated below, we  assume that $p$, $q$, and $\chi$ are as in Theorem \ref{1tm}
\begin{lemma} [Integrability]
 \label{New1lm}
For every $\lambda>0$, there exist positive constants $\mathit{\Upsilon}_i(\lambda)$, $i=1,2,3$, such that for every $\mathit{\Delta}\Subset \mathbb{Z}^d$, and for any $J\in \mathcal{J}_q$ and $\xi\in \mathcal{S}_p$, the following holds
\begin{eqnarray}
 \label{New1}
& & \int_{\mathbb{R}^{\mathbb{Z}^d}} \exp\left(\lambda \| \sigma\|_p^p \right) \pi_{\mathit{\Delta}} (d \sigma |J , \xi)\\[.2cm]& & \qquad \qquad \leq  \exp\left( \mathit{\Upsilon}_1(\lambda) + \mathit{\Upsilon}_2(\lambda) \|J\|_q^q + \mathit{\Upsilon}_3(\lambda) \|\xi_{\mathit{\Delta}^c}\|^p_p \right). \nonumber
\end{eqnarray}
\end{lemma}
\begin{corollary}
  \label{1co}
For every fixed $\xi \in \mathcal{S}_{p} $ and $R>0$, the family
\[
\{\pi_{\mathit{\Delta}}( \cdot|J ,\xi) \ : \ \mathit{\Delta}\Subset \mathbb{Z}^d,  \ \|J\|_q \leq R\} \subset \mathcal{P}(\mathbb{R}^{\mathbb{Z}^d})
\]
is relatively compact.
\end{corollary}
{\it Proof:}
By (\ref{J14aa}), for any positive $R$ and $\lambda$, and for every $\mu \in \mathcal{G}_p(J)$ with $J \in B_q(R)$, one has
\begin{equation}
  \label{j14b}
\int_{\mathbb{R}^{\mathbb{Z}^d}} \exp\left(\lambda \|\sigma\|^p_p  \right)\mu(d\sigma) \leq \exp\left( \mathit{\Upsilon}_1 (\lambda) + \mathit{\Upsilon}_2 (\lambda)R^q\right).
\end{equation}
Obviously, for any $p>0$, the balls $B_p(r) = \{\sigma : \|\sigma\|_{p}\leq r\}$, $r>0$, are compact in the product topology of
$\mathbb{R}^{\mathbb{Z}^d}$. Then the proof follows from (\ref{j14b}) by Prohorov's theorem.$\square$ \vskip.1cm \noindent
Let us consider the map
\begin{equation}
  \label{j32}
C_{\rm b} (\mathbb{R}^{\mathbb{Z}^d}) \ni f \mapsto \int_{\mathbb{R}^{\mathbb{Z}^d}} f(\sigma) \pi_{\mathit{\Delta}} ( d\sigma |J,\cdot).
\end{equation}
\begin{lemma}[Feller property]
  \label{4lm}
For every $\mathit{\Delta}\Subset \mathbb{Z}^d$ and $J\in \mathcal{J}_q $, the image of (\ref{j32}) is in $C_{\rm b} (\mathbb{R}^{\mathbb{Z}^d})$.
\end{lemma}
The proof of this lemma is quite standard. The boundedness of the right-hand side of (\ref{j32}) is immediate. The continuity follows by Lebesgue's dominated convergence theorem from the continuity of the function (\ref{j3}). For more details we refer the reader to the proof of Lemma 2.10 in \cite{[KP]}.
\begin{lemma}[Lipschitz continuity]
 \label{New2lm}
For every $\mathit{\Delta} \Subset \mathbb{Z}^d$ and any $R>0$, there exist positive constants $\mathit{\Theta}_i(\mathit{\Delta}, R)$, $i=1,2$, such that for every $J, J' \in B_q(R)$, any $f\in C_{\rm b} (\mathbb{R}^{\mathbb{Z}^d})$ and any $\xi\in \mathcal{S}_p$, the following holds
\begin{eqnarray} \label{New2}
& & \left\vert \int_{\mathbb{R}^{\mathbb{Z}^d}} f(\sigma) \pi_{\mathit{\Delta}} ( d\sigma |J,\xi) - \int_{\mathbb{R}^{\mathbb{Z}^d}} f(\sigma) \pi_{\mathit{\Delta}} ( d\sigma |J',\xi) \right\vert \\[.2cm]
& & \qquad \qquad \leq \|J - J'\|_q \|f\|_{\infty} \left( \mathit{\Theta}_1 (\mathit{\Delta}, R) + \mathit{\Theta}_2 (\mathit{\Delta}, R) \|\xi\|_p^p\right). \nonumber
\end{eqnarray}
\end{lemma}

\subsection{The proof of Theorem \ref{1tm}}
We first prove that $\mathcal{G}_p (J)$ is non-void. Let us fix some $\xi \in \mathcal{S}_p $. For every $\mathit{\Delta}\Subset \mathbb{Z}^d$, the measure
$\pi_{\mathit{\Delta}}( \cdot|J,\xi)$ is supported on the set $\{\sigma = \sigma_{\mathit{\Delta}} \times \xi_{\mathit{\Delta}^c} : \sigma_{\mathit{\Delta}} \in \mathbb{R}^{|\mathit{\Delta}|}\}$, see (\ref{j5}). This yields
 \begin{equation}
 \label{j33}
\pi_{\mathit{\Delta}}\left(\mathcal{S}_p|J,\xi\right) =1.
 \end{equation}
By Corollary \ref{1co}, there exists a cofinal sequence $\{\mathit{\Delta}_n \}_{n\in \mathbb{N}}$ such that the sequence of measures
$\{\pi_{\mathit{\Delta}_n}( \cdot|J,\xi)\}_{n\in \mathbb{N}}$ converges to a certain $\mu \in \mathcal{P}(\mathbb{R}^{\mathbb{Z}^d})$. Let us show that this $\mu$ solves the DLR equation (\ref{jj0}), that is,
\begin{equation}
 \label{j33a}
\int_{\mathbb{R}^{\mathbb{Z}^d}} \left\{\int_{\mathbb{R}^{\mathbb{Z}^d}} f(\sigma) \pi_{\mathit{\Delta}} ( d\sigma|J,\eta)\right\} \mu (d\eta) = \int_{\mathbb{R}^{\mathbb{Z}^d}} f(\sigma) \mu (d\eta),
\end{equation}
holding for all $\mathit{\Delta} \Subset \mathbb{Z}^d$ and all $f\in C_{\rm b}(\mathbb{R}^{\mathbb{Z}^d})$.
For any $\mathit{\Delta} \Subset \mathbb{Z}^d$, one finds $m\in \mathbb{N}$ such that $\mathit{\Delta} \subset \mathit{\Delta}_n$ for all $n\geq m$.
For such $n$ and any $f\in C_{\rm b}(\mathbb{R}^{\mathbb{Z}^d})$, by (\ref{i2}) we get
\begin{equation}
 \label{j34}
\int_{\mathbb{R}^{\mathbb{Z}^d}} \left\{\int_{\mathbb{R}^{\mathbb{Z}^d}} f(\sigma) \pi_{\mathit{\Delta}} ( d\sigma|J,\eta)\right\} \pi_{\mathit{\Delta}_n} ( d\eta|J,\xi) = \int_{\mathbb{R}^{\mathbb{Z}^d}} f(\sigma) \pi_{\mathit{\Delta}_n} ( d\sigma|J,\xi).
\end{equation}
Now we pass here to the limit $n\rightarrow+\infty$ and obtain (\ref{j33a}) by Lemma \ref{4lm}. To prove that this $\mu$ is supported on $\mathcal{S}_p$, let us show that it obeys the estimate (\ref{j14b}). For $\lambda >0$ and $N\in \mathbb{N}$, we set
\begin{equation}
 \label{j35}
F_N (\sigma) = \exp \left(\lambda \min\{ \|\sigma\|^p_p ; N\} \right), \quad \ \sigma\in \mathbb{R}^{\mathbb{Z}^d}.
\end{equation}
Such functions are lower semi-continuous. Then in view of (\ref{New1}) and of the fact that $\xi\in \mathcal{S}_p$,
by Fatou's lemma
we have
\begin{eqnarray*}
\int_{\mathbb{R}^{\mathbb{Z}^d}} F_N(\sigma) \mu(d \sigma ) & \leq & \lim_{n\rightarrow +\infty}\int_{\mathbb{R}^{\mathbb{Z}^d}} F_N(\sigma) \pi_{\mathit{\Delta}_n} ( d\eta|J, \xi)\\[.2cm] &\leq & \exp\left(\mathit{\Upsilon}_1(\lambda) + \mathit{\Upsilon}_2(\lambda) \|J\|_q^q \right).
\end{eqnarray*}
Thereafter, by B. Levi's monotone convergence theorem we obtain that: (a) the limiting measure is in $\mathcal{G}_p (J)$; (b) each such a measure obeys the estimate (\ref{J14aa}) with the constants as in Lemma \ref{New1lm}.
Now to complete the proof we have to show that: (c) the estimate (\ref{j14b}) holds for all $\mu\in \mathcal{G}_p (J)$; (d) the set $\mathcal{G}_p (J)$ is compact. Let $\mu$ be an arbitrary element of $\mathcal{G}_p (J)$. By (\ref{jj0}), Fatou's lemma and (\ref{New1}),  we get
\begin{eqnarray*}
& & \int_{\mathbb{R}^{\mathbb{Z}^d}} F_N (\sigma)  \mu(d \sigma ) = \limsup_{\mathit{\Delta}\nearrow \mathbb{Z}^d} \int_{\mathbb{R}^{\mathbb{Z}^d}}\left\{\int_{\mathbb{R}^{\mathbb{Z}^d}} F_N (\sigma)  \pi_{\mathit{\Delta}} ( d\sigma|J, \xi)\right\}\mu( d \xi)\\[.2cm] & & \qquad \leq \int_{\mathbb{R}^{\mathbb{Z}^d}}\left\{\limsup_{\mathit{\Delta}\nearrow \mathbb{Z}^d}\int_{\mathbb{R}^{\mathbb{Z}^d}} F_N (\sigma)  \pi_{\mathit{\Delta}} ( d\sigma|J,\xi)\right\}\mu( d \xi)\\[.2cm] & & \qquad \leq  \int_{\mathbb{R}^{\mathbb{Z}^d}}\left\{\limsup_{\mathit{\Delta}\nearrow \mathbb{Z}^d}\int_{\mathbb{R}^{\mathbb{Z}^d}}
\exp \left(\lambda \|\sigma\|^p_p  \right)\pi_{\mathit{\Delta}} (J, d\sigma|\xi)\right\}\mu( d \xi)\\[.2cm] & & \qquad \leq
\exp\left(\mathit{\Upsilon}_1 (\lambda) + \mathit{\Upsilon}_2 (\lambda)\|J\|_q^q \right).
\end{eqnarray*}
Then we again apply B. Levi's theorem and obtain (\ref{J14aa}). In view of this estimate, by Prokhorov's theorem the set $\mathcal{G}_p (J)$ is relatively compact. All of its accumulation points clearly solve the DLR equation (\ref{j33a}); hence,  $\mathcal{G}_p (J)$ is compact.$\square$

\subsection{The proof of Theorem \ref{2tm}}

We recall that  $\mathcal{P}(\mathbb{R}^{\mathbb{Z}^d})$ is a Polish space. The  latter fact is important for the following reason. By the fundamental theorem of the set-valued analysis\footnote{See Theorem 8.1.3 in \cite{[Aubin]}, page 308.}, a map from a measurable space to non-void closed subsets of a Polish space admits a measurable selection if it is measurable. By Theorem \ref{1tm}, the images of the map $\mathcal{J}_q \ni J \mapsto \mathcal{G}(J) \subset \mathcal{P}(\mathbb{R}^{\mathbb{Z}^d})$ are compact and hence closed.
According to Definition 8.1.1 in \cite{[Aubin]}, page 307, the map $\mathcal{J}_q \ni J \mapsto \mathcal{G}(J)$ is {\it measurable} if for every open $A\subset \mathcal{P}(\mathbb{R}^{\mathbb{Z}^d})$, the set
\begin{equation}
 \label{Jj}
\mathcal{G}^{-1}_{p}(A) = \{ J \in \mathcal{J}_q  \ : \ \mathcal{G}_{p}(J) \cap A \neq \emptyset\}
\end{equation}
is measurable. Since the probability space $(\mathcal{J}_q, \mathcal{B}(\mathcal{J}_q), \nu)$ is complete, the measurability in question can be obtained from the fact that the map is upper semi-continuous, see Proposition 8.2.1 in \cite{[Aubin]}, page 311. In our case, the latter means that the set
$\mathcal{G}^{-1}_{p}(A)$ is closed whenever $A$ is closed, see Proposition 1.4.4 in \cite{[Aubin]}, page 40.
Thus, to prove the existence of measurable selections  we have only to show the upper semi-continuity just mentioned.
To this end it is enough to demonstrate that for any Cauchy sequence $\{J_n\}_{n\in \mathbb{N} }\subset \mathcal{G}^{-1}_{p}(A)$, its limit $J$ is also in $\mathcal{G}^{-1}_{p}(A)$. Let $R>0$ be such that the sequence, as well as its limit, are contained in the ball $B_q(R)$. For each $n\in \mathbb{N}$, we take  $\mu_n \in \mathcal{G}_p (J_n)\cap A$. Then all the elements of the sequence $\{\mu_n\}_{n\in \mathbb{N}}\subset A$ obey the estimate (\ref{j14b}). By Prohorov's theorem, this yields that $\{\mu_n\}_{n\in \mathbb{N}}$ is relatively compact in $\mathcal{P}(\mathbb{R}^{\mathbb{Z}^d})$. Each of its accumulation points $\mu$ obeys (\ref{j14b}), see the proof of Theorem \ref{1tm}, and hence is supported on $\mathcal{S}_p$. Therefore, each such $\mu$
is in  $\mathcal{P}_{\rm temp}\cap A$ as $A$ is closed.
This means that $\mathcal{G}_p (J) \cap A \neq \emptyset$ if
this $\mu$ solves (\ref{jj0}) with any $f\in C_{\rm b}(\mathbb{R}^{\mathbb{Z}^d})$ and any $\mathit{\Delta}\Subset \mathbb{Z}^d$. For fixed such $f$ and $\mathit{\Delta}$, we set, c.f. (\ref{jj0}) and (\ref{jj1}),
\begin{equation}
  \label{Jk}
 \delta_{\mathit{\Delta}} (f) = \left\vert \mu(f) - \int_{\mathbb{R}^{\mathbb{Z}^d}}\pi_{\mathit{\Delta}} (f|J, \xi)\mu(d\xi) \right\vert
\end{equation}
Let $\{n_k\}_{k\in \mathbb{N}}$ be such that $\mu_{n_k} \rightarrow \mu$ in $\mathcal{P}(\mathbb{R}^{\mathbb{Z}^d})$. Then for any $n_k$, we have
\begin{equation}
 \label{Jk1}
\delta_{\mathit{\Delta}} (f) \leq \left\vert \mu(f) - \mu_{n_k} (f)  \right\vert +
\left\vert \mu(g) - \mu_{n_k} (g)  \right\vert + \theta_{n_k} (\mathit{\Delta}, f),
\end{equation}
where $g= \pi_{\mathit{\Delta}} (f| J, \cdot)$ and
\begin{equation} \label{Jk2}
 \theta_{n_k} (\mathit{\Delta}, f) = \int_{\mathbb{R}^{\mathbb{Z}^d}} \left\vert  \pi_{\mathit{\Delta}} (f| J, \xi) - \pi_{\mathit{\Delta}} (f| J_{n_k}, \xi)\right\vert \mu_{n_k} (d\xi).
\end{equation}
The first two terms in (\ref{Jk1}) can be made arbitrarily small, see Lemma \ref{4lm}. Let us show that this is true also for the third term.
By Jensen's inequality, we have from (\ref{j14b})  that
\[
\int_{\mathbb{R}^{\mathbb{Z}^d}} \|\xi\|_p^p \mu_{n_k} (d\xi) \leq \left(\mathit{\Upsilon}_1(\lambda) + \mathit{\Upsilon}_2(\lambda)R^d \right)/\lambda,
\]
which holds for any fixed $\lambda>0$.
Then we apply in (\ref{Jk2}) Lemma \ref{New2lm} and the latter estimate, and arrive at
\begin{eqnarray*}
\theta_{n_k} (\mathit{\Delta}, f) & \leq & \|J - J_{n_k} \|_q \|f\|_{\infty} \\[.2cm] & \times & \left[\mathit{\Theta}_1 (\mathit{\Delta}, R) + \mathit{\Theta}_2 (\mathit{\Delta}, R)\left( \mathit{\Upsilon}_1(\lambda) + \mathit{\Upsilon}_2(\lambda)R^d \right)/\lambda \right],
\end{eqnarray*}
which completes the proof of the upper semi-continuity of the map $\mathcal{J}_q \ni J \mapsto \mathcal{G}_p (J )$ and hence of the whole statement.$\square$

\subsection{The proof of Theorems \ref{3tm} and {\ref{4tm}}}

In the measure-theoretic context,  the statements of these theorems are about the almost sure convergence of the sequences of conditional distributions 
which  should be obtained from the weak convergence of the corresponding measures, and some additional facts.  
Clearly, it would be too much to expect that it holds directly for the sequences. That is why we consider the Ces\`{a}ro-like means (\ref{Co1}) and (\ref{Co2}) to which we apply 
 the Koml{\'o}s theorem \cite{[Komlos]} presented here in the form taken from \cite{[Balder]}.
\begin{proposition}[Koml{\'o}s theorem]
\label{Kompn}
 Let $(\Omega, \mathcal{F}, \mu)$ be a finite measure space and $L^1(\mu)$ be the space of integrable real-valued functions.
Suppose also that a sequence $\{x_n\}_{n\in \mathbb{N}} \subset L^1 (\mu)$ is such that
\[
\sup_n \int_{\Omega} |x_n(\omega)|  \mu(d \omega) < +\infty.
\]
Then there exists $y\in L^1 (\mu)$ and a subsequence $\{x_{n_k}\}$ of  $\{x_n\}$ such that for every further
subsequence $\{x_{n_{k_m}}\}$ of  $\{x_{n_k}\}$, the following holds
\begin{equation}
 \label{komlos}
\frac{1}{M} \sum_{m=1}^{M} x_{n_{k_m}} (\omega) \rightarrow y(\omega), \ \ M\rightarrow +\infty, \quad {\rm for} \ \ \mu-{\rm almost} \ {\rm all} \ \omega \in \Omega.
\end{equation}
\end{proposition}
{\it Proof of Theorem \ref{3tm}:} We fix some $\xi \in \mathcal{J}_q$ and show that there exists a measurable map
$\mathcal{J}_q \ni J \mapsto \mu^\xi (J) \in \mathcal{P}(\mathbb{R}^{\mathbb{Z}^d})$ such that the sequence of the averaged kernels $\{\pi_{\mathcal{D},N}\}_{N\in \mathbb{N}}$ defined in (\ref{Co1}) converges to
$ \mu^\xi (J)$ for some cofinal sequence $\mathcal{D}$ and $\nu$-almost all $J$. As in the proof of Theorem \ref{1tm}
this will imply that $\mu^\xi (J) \in \mathcal{G}_p (J)$. For the mentioned $\xi$ and $\mathit{\Delta}\Subset \mathbb{Z}^d$, we define
\begin{equation}
  \label{dfg1}
\vartheta^\xi_{\mathit{\Delta}} (d \sigma , d J) = \pi_{\mathit{\Delta}} (d \sigma|J, \xi) \nu(dJ),
\end{equation}
which is a probability measure on $\mathbb{R}^{\mathbb{Z}^d} \times \mathcal{J}_q$.
In the product topology, $\mathbb{R}^{\mathbb{Z}^d} \times\mathcal{J}_q$ is a Polish space. By $\mathcal{P}( \mathbb{R}^{\mathbb{Z}^d} \times \mathcal{J}_q)$ we denote the space of all probability measures defined thereon, equipped with the usual weak topology. Let us show that the family $\{\vartheta^\xi_{\mathit{\Delta}}\}_{\mathit{\Delta}\Subset \mathbb{Z}^d}$ is relatively compact.
By construction, each $\nu^\xi_{\mathit{\Delta}}$ is supported on $\mathcal{S}_p \times \mathcal{J}_q$. For every $r>0$, the ball $B_p(r) = \{ \sigma \in \mathcal{S}_p : \|\sigma\|_p \leq r\}$ is compact in
$\mathbb{R}^{\mathbb{Z}^d}$. By (\ref{New1}), we readily get
\begin{eqnarray}
  \label{dfg2}
& &  \pi_{\mathit{\Delta}} \left( B_p(r)| J, \xi  \right) \\[.2cm] & & \qquad \geq 1 - \exp\left(- \lambda r^p + \mathit{\Upsilon}_1 (\lambda) +    \mathit{\Upsilon}_2 (\lambda) \|J\|_q^q + \mathit{\Upsilon}_3 (\lambda) \|\xi\|_p^p \right). \nonumber
\end{eqnarray}
Given $\varepsilon > 0$, let $\mathcal{J}_q^\varepsilon \subset \mathcal{J}_q$ be compact and such that
$\nu (\mathcal{J}_q^\varepsilon) \geq 1 - \varepsilon/2$. Clearly, this
$\mathcal{J}_q^\varepsilon$ is contained in the ball $B_q(R_\varepsilon)$ for a sufficiently big
$R_\varepsilon$. Then we pick $r_\varepsilon$ such that
\[
\exp\left(- \lambda r^p_\varepsilon + \mathit{\Upsilon}_1 (\lambda) +    \mathit{\Upsilon}_2 (\lambda) R_\varepsilon^q + \mathit{\Upsilon}_3 (\lambda) \|\xi\|_p^p \right) < \varepsilon /2,
\]
for some fixed $\lambda >0$. Then by (\ref{dfg2}) we immediately obtain that
\[
 \vartheta^\xi_{\mathit{\Delta}} \left(B_p (r_\varepsilon) \times \mathcal{J}_q^\varepsilon \right) \geq 1 - \varepsilon,
\]
which holds for any $\mathit{\Delta} \Subset \mathbb{Z}^d$. Thus, by Prohorov's theorem the family $\{\vartheta^\xi_{\mathit{\Delta}}\}_{\mathit{\Delta}\Subset \mathbb{Z}^d}$ is relatively compact and hence has accumulation points. Let $\vartheta^\xi$ be any of them. In view of the mentioned convergence, for any $g \in C_{\rm b} (\mathcal{J}_q)$ we have that
\begin{equation} \label{Co3}
\int_{\mathbb{R}^{\mathbb{Z}^d} \times \mathcal{J}_q} g(J ) \vartheta^\xi (d \sigma , dJ) = \int_{ \mathcal{J}_q} g(J ) \nu (d J),
\end{equation}
which yields that the projection of $\vartheta^\xi$ onto
$\mathcal{J}_q$ is $\nu$. Since $\mathbb{R}^{\mathbb{Z}^d} \times \mathcal{J}_q$ is a Polish space,
by (\ref{Co3}) we can disintegrate\footnote{See Theorem 8.1 on page 147 in \cite{[Part]}.}
\begin{equation}
  \label{dfg3} \vartheta^\xi (d \sigma , dJ) = \vartheta^\xi (d \sigma |J) \nu(d J),
\end{equation}
where $\vartheta^\xi (d \sigma |J)$ is a regular conditional distribution.
Let $\mathcal{D}_0 = \{\mathit{\Delta}_n\}_{n\in \mathbb{N}}$ be the cofinal sequence along which the sequence $\{\vartheta^\xi_{\mathit{\Delta}_n}\}_{n\in \mathbb{N}}$ converges to this
$\vartheta^\xi$. In view of (\ref{dfg1}) and (\ref{dfg3}), this means that for every $f \in C_{\rm b} (\mathbb{R}^{\mathbb{Z}^d})$ and $g\in C_{\rm b}(\mathcal{J}_q)$,
\begin{equation}
  \label{dfg4}
\int_{\mathcal{J}_q}g(J) \pi_{\mathit{\Delta}_n} (f |J, \xi) \nu(d J) \rightarrow \int_{\mathcal{J}_q}g(J)\vartheta^\xi (f|J) \nu(d J), \ \  n\rightarrow +\infty.
\end{equation}
From the latter convergence one cannot get that $\pi_{\mathit{\Delta}_n} (f|J,\xi) \rightarrow \vartheta^\xi(f|J)$ for $\nu$-almost all $J$. However, (\ref{dfg4}) can be of use if we apply the Koml\'os theorem.
As $\pi_{\mathit{\Delta}} (\cdot |J,\xi)$ is a probability measure for all $J\in \mathcal{J}_q$ and $\xi \in \mathcal{S}_p$, for any $f \in C_{\rm b} (\mathbb{R}^{\mathbb{Z}^d})$ we have that
\begin{equation} \label{dfg50} \pi_{\mathit{\Delta}_n} (f |J, \xi) \leq \|f\|_{\infty}.
\end{equation}
Since the  topology of $\mathcal{P} (\mathbb{R}^{\mathbb{Z}^d})$ is metrizable, there exists a family\footnote{See page 19 in \cite{[Borkar]}.} $\{f_i\}_{i\in \mathbb{N}} \in C_{\rm b} (\mathbb{R}^{\mathbb{Z}^d})$ such that the convergence of a sequence $\{\mu_k\}_{k\in \mathbb{N}}\subset  \mathcal{P} (\mathbb{R}^{\mathbb{Z}^d})$ to a certain $\mu$ holds if
$\mu_k (f_i) \rightarrow \mu(f_i)$ for every $i\in \mathbb{N}$, c.f. (\ref{jj1}). By the Koml\'os theorem, from (\ref{dfg4}) and (\ref{dfg50}) we have that there exists a sequence $\mathcal{D}_1 \subset \mathcal{D}_0$ such that for any
$\mathcal{D}_1' \subset \mathcal{D}_1$, we have that
\begin{equation}
  \label{dfg51}
  \pi_{\mathcal{D}_1',N} (f_1|J, \xi) \rightarrow \vartheta^\xi (f_1|J), \quad  N\rightarrow +\infty,
\end{equation}
which holds for all $J\in A'_1 \subset \mathcal{J}_q$, such that $\nu
(A'_1)=1$. We take any such $\mathcal{D}_1'$ and apply to it the same
arguments as to $\mathcal{D}_0$, which yields that for some
$\mathcal{D}_2 \subset \mathcal{D}_1'$ and for any $\mathcal{D}'_2
\subset \mathcal{D}_2$, we have that alone with (\ref{dfg51}) the
same convergence holds also for the sequence $\{
\pi_{\mathcal{D}_2',N} (f_2|J, \xi)\}_{N\in \mathbb{N}}$ with all
$J\in A'_2\subset \mathcal{J}_q$, such that $\nu (A'_2)=1$. Continuing this procedure we obtain the sequence
$\{\mathcal{D}'_i\}_{i \in \mathbb{N}}$, $\mathcal{D}'_{i+1} \subset
\mathcal{D}'_i$, and the family $\{A'_i\}_{i \in \mathbb{N}}$. Set $A
= \cap_{i} A'_i$ and let  $\mathcal{D}$ be the diagonal sequence
which one obtains by taking the first element of $\mathcal{D}'_1$,
the second element of $\mathcal{D}'_2$, and so on. Then $\nu(A) = 1$
and the convergence
\[
\pi_{\mathcal{D},N} (f_i|J, \xi) \rightarrow \vartheta^\xi (f_i|J), \quad  N\rightarrow +\infty,
\]
holds for all $i\in \mathbb{N}$ and all $J\in A$. This yields the
convergence of the sequence $\{\pi_{\mathcal{D},N} (\cdot |J,
\xi)\}_{N \in \mathbb{N}}$ to the measure $\vartheta^\xi(\cdot |J)$
which holds for all $J\in A$, $\nu(A) =1$. As in the proof of
Theorem \ref{1tm}, this convergence implies that
$\vartheta^\xi(\cdot |J)\in \mathcal{G}_p (J)$ which holds for
$\nu$-almost all $J$. Then we set $\mu^\xi (J) = \vartheta^\xi(\cdot
|J)$, c.f. (\ref{pres2}). $\square$ \vskip.2cm \noindent {\it Proof of Theorem
\ref{4tm}:} For $\xi\in \mathcal{S}_p$ and $\mathit{\Delta}\Subset
\mathbb{Z}^d$, we set, c.f. (\ref{Co2}),
\begin{equation}
  \label{dfg6}
\mathfrak{t}^\xi_{\mathit{\Delta}} (d \mu , dJ) = \mathfrak{d}^\xi_{\mathit{\Delta}}(J) (d \mu ) \nu (d J),
\end{equation}
which is a probability measure on the product
$\mathcal{P}(\mathbb{R}^{\mathbb{Z}^d})\times \mathcal{J}_q$.
Similar as above, we equip this set with the product topology and
hence turn it into a Polish space.  Let us show that the family
$\{\mathfrak{t}^\xi_{\mathit{\Delta}}\}_{\mathit{\Delta}\Subset
\mathbb{Z}^d}$ is tight. By Corollary \ref{1co}, for every $R>0$ the
closure in $\mathcal{P}(\mathbb{R}^{\mathbb{Z}^d})$ of the family of
kernels, that is, the set
\[
\mathit{\Pi}^\xi (R) = \overline{\{\pi_{\mathit{\Delta}} (\cdot|J,
\xi) \ : \ \mathit{\Delta} \Subset \mathbb{Z}^d, \ J \in B_q(R)\}}
\subset \mathcal{P}(\mathbb{R}^{\mathbb{Z}^d})
\]
is compact. Clearly $\mathfrak{d}^\xi_{\mathit{\Delta}}(J)
(\mathit{\Pi}^\xi(R)) =1$ for every $\mathit{\Delta} \Subset
\mathbb{Z}^d$ and $R>0$. Let $\mathcal{J}^\varepsilon$ be compact
and such that $\nu(\mathcal{J}^\varepsilon) \geq 1 -\varepsilon$.
Then   $\mathcal{J}^\varepsilon \subset B_q(R_\varepsilon)$ for
sufficiently big $R_\varepsilon$. Therefore,
\[
\mathfrak{t}^\xi_{\mathit{\Delta}} \left( \mathit{\Pi}^\xi
(R_\varepsilon )\times \mathcal{J}^\varepsilon \right) \geq 1 -
\varepsilon,
\]
which yields the tightness and hence the relative weak compactness
of the family
$\{\mathfrak{t}^\xi_{\mathit{\Delta}}\}_{\mathit{\Delta}\Subset
\mathbb{Z}^d}$. Let $\mathfrak{t}^\xi$ be any of its accumulation
points. As in the proof of Theorem \ref{3tm}, one shows that the
projection of $\mathfrak{t}^\xi$ onto $\mathcal{J}_q$ is $\nu$. This allows us to disintegrate,
 c.f. (\ref{dfg3}),
\begin{equation}
  \label{dfg7}
\mathfrak{t}^\xi (d\mu , dJ) = \mathfrak{t}^\xi (d\mu|J) \nu (d J).
\end{equation}
Let $\{\mathit{\Delta}_n\}_{n\in \mathbb{N}}$ be the cofinal
sequence along which the sequence
$\{\mathfrak{t}^\xi_{\mathit{\Delta}_n}\}_{n\in \mathbb{N}}$
converges to $\mathfrak{t}^\xi$. One  can show that
\begin{equation}
  \label{dfg8}
\int_{\mathcal{J}_q} g(J) \mathfrak{d}^\xi_{\mathit{\Delta}_n}(J)
(F) \nu(dJ) \rightarrow \int_{\mathcal{J}_q} g(J)
\mathfrak{t}^\xi(F|J)
 \nu(dJ), \quad n \rightarrow +\infty,
\end{equation}
which holds for all $g\in C_{\rm b} (\mathcal{J}_q)$ and $F\in
C_{\rm b} (\mathcal{P} (\mathbb{R}^{\mathbb{Z}^d}))$. As in the
proof of Theorem \ref{3tm}, by means of the Koml{\'o}s theorem we
show that (\ref{dfg8}) implies the existence of a cofinal sequence
$\mathcal{D} = \{\mathit{\Delta}_{m}\}_{m\in \mathbb{N}}$ such that
\begin{equation}
  \label{dfg9}
\frac{1}{M} \sum_{m=1}^M \mathfrak{d}^\xi_{\mathit{\Delta}_m}(J)
\rightarrow \mathfrak{t}^\xi(\cdot|J), \quad M\rightarrow +\infty,
\end{equation}
where the convergence is in the space $\mathfrak{P}$ and holds for
$\nu$-almost all $J$. For every $f\in C_{\rm b}
(\mathbb{R}^{\mathbb{Z}^d})$, the evaluation map
\[
\mathcal{P}(\mathbb{R}^{\mathbb{Z}^d}) \ni \mu \mapsto \mu(f) \in
\mathbb{R}
\]
is clearly in $C_{\rm b} (\mathcal{P} (\mathbb{R}^{\mathbb{Z}^d}))$.
Hence, by (\ref{dfg9}) we have that
\begin{eqnarray}
  \label{dfg10}
& & \frac{1}{M} \sum_{m=1}^M
\int_{\mathcal{P}(\mathbb{R}^{\mathbb{Z}^d})} \mu(f)
\mathfrak{d}^\xi_{\mathit{\Delta}_m}(J) (d\mu)  = \frac{1}{M}
\sum_{m=1}^M \pi_{\mathit{\Delta}_m} (f |J, \xi)
 \\[.2cm]
& & =  \int_{\mathcal{P}(\mathbb{R}^{\mathbb{Z}^d})} \mu(f)
\mathfrak{d}_{\mathcal{D}, M}^\xi (J) (d\mu) \rightarrow
\mathfrak{t}^\xi(\mu(f)|J), \quad M\rightarrow +\infty, \nonumber
\end{eqnarray}
which holds for all $f\in C_{\rm b} (\mathbb{R}^{\mathbb{Z}^d})$.
From the latter convergence we see that
\[
\mu^\xi (J) \ \stackrel{\rm def}{=} \
\int_{\mathcal{P}(\mathbb{R}^{\mathbb{Z}^d})} \mu \ \mathfrak{t}^\xi
(d \mu|J)
\]
is a random tempered Gibbs measure. Then
$\mathfrak{m}^\xi(J)\stackrel{\rm def}{=} \mathfrak{t}^\xi
(\cdot|J)$ is an Aizenman-Wehr metastate, which completes the proof.
$\square$

\subsection{The proof of Theorems \ref{5tm} and \ref{6tm}}

The proof of the both theorems is based on a stronger version of the estimate (\ref{J14aa}), which
we obtain under the condition of the uniform integrability assumed in (\ref{pres3}). We formulate
it in the following lemma the proof of which is given in Section \ref{5SE}
\begin{lemma}
 \label{integrlm}
Let $\nu$ be as in Theorem \ref{5tm}. Then there exists $c_\nu>0$ such that for every
$x\in \mathbb{Z}^d$ and for any $\mathit{\Delta}\Subset \mathbb{Z}^d$ (resp. for any random Gibbs measure $\mu(J)$), the estimate (\ref{integ}) (resp. (\ref{integG})) hold
\begin{eqnarray}
 \label{integ}
& & \int_{\mathbb{R}^{\mathbb{Z}^d} \times \mathcal{J}_q} |\sigma (x)|^p \pi_{\mathit{\Delta}} (d \sigma |J,0) \nu(dJ) \leq c_{\nu},\\[.2cm]
\label{integG}
& & \int_{\mathbb{R}^{\mathbb{Z}^d} \times \mathcal{J}_q} |\sigma (x)|^p \vartheta (d \sigma , d J) \leq c_{\nu},
\end{eqnarray}
where $\vartheta$ is as in (\ref{pres2}).
\end{lemma}
Note that these estimates are uniform in $x$. By Jensen's inequality, directly from (\ref{J14aa}) we can get that ${\rm LHS}(\ref{integG}) \leq {\rm const}/w(x)$, c.f. (\ref{exa}), which is not enough for proving the theorem.
\vskip.1cm \noindent
{\it Proof of Theorem \ref{5tm}:}
For $p$ and $q$ as in Theorem \ref{1tm} and for positive $a,b,c, \varkappa$,  by means of the Young inequality  one can prove that
\begin{equation} \label{fz}
a b c \leq \varkappa (b^p + c^p) + (p-2) p^{-p/(p-2)} \varkappa^{-2/(p-2)} a^{p/(p-2)}.
\end{equation}
We use (\ref{fz}) with $\varkappa =1$ to estimate the interaction energy (\ref{j4})
\begin{eqnarray}
 \label{een}
\left\vert H_{\mathit{\Delta}} (\sigma_{\mathit{\Delta}}|J,\xi) \right\vert & \leq & 2 d \sum_{x\in \mathit{\Delta}}|\sigma(x) |^p + 2 d \sum_{y\in \partial \mathit{\Delta}}|\xi (y)|^p\\[.2cm]& + & \frac{1}{2} \sum_{x\in \mathit{\Delta}} \sum_{y\sim x} |J_{xy}|^q, \nonumber
\end{eqnarray}
where $\partial \mathit{\Delta} = \{ y \in \mathit{\Delta}^c : \exists x \in \mathit{\Delta} \ y \sim x\}$. Here we have taken into account that $(p-2)/p^{p/(p-2)} = 2 (q-1)^{q-1}/ (2q)^q$ and
\begin{equation} \label{lipq}
(q-1)^{q-1}/{(2q)^q} \leq 1/2, \qquad {\rm for} \quad   q \geq 1.
\end{equation}
Then we employ (\ref{een}) in (\ref{pres}) and obtain
\begin{eqnarray}
 \label{enn1}
\left\vert p_{\mathit{\Delta}} (J,\xi) \right\vert & \leq & \frac{2d}{|\mathit{\Delta}|}\sum_{y\in \partial \mathit{\Delta}}|\xi (y)|^p + \frac{1}{2|\mathit{\Delta}|} \sum_{x\in \mathit{\Delta}} \sum_{y\sim x} |J_{xy}|^q\\[.2cm] & + & \max\{ \log C_{+} ( 2d) ; - \log C_{-} (2d)\} , \nonumber
\end{eqnarray}
where the constants $C_{\pm} ( 2d)$ are the same as in (\ref{j13}).
By (\ref{enn1}), (\ref{pres3}), and (\ref{integG}) we obtain that
\begin{eqnarray*}
 \sup_{\mathit{\Delta} \Subset \mathbb{Z}^d} \int_{ \mathcal{J}_q} \left\vert
  \bar{p}^\mu_{\mathit{\Delta}} (J) \right\vert \nu(dJ) & < & \infty\\[.2cm]\sup_{\mathit{\Delta} \Subset \mathbb{Z}^d} \int_{\mathbb{R}^{\mathbb{Z}^d }\times \mathcal{J}_q} \left\vert p_{\mathit{\Delta}} (J,\xi) \right\vert \vartheta(d\xi , dJ) & < & \infty.
\end{eqnarray*}
Then the proof of the both statements follows by the Koml{\'o}s theorem. $\square$ \vskip.1cm \noindent
{\it Proof of Theorem \ref{6tm}:}
To prove (\ref{pres5}) we use the corresponding arguments of \cite{[Cont],[Cont1]}.
Given $\mathit{\Delta} \Subset \mathbb{Z}^d$, we take an arbitrary  $\langle \bar{x}, \bar{y} \rangle \in
{\sf E}_{\mathit{\Delta}}$ and set
\begin{equation}
 \label{inte}
-\bar{H}_\mathit{\Delta} (\sigma_{\mathit{\Delta}}|J,\xi) = - {H}_\mathit{\Delta} (\sigma_{\mathit{\Delta}}|J,\xi) -
J_{\bar{x} \bar{y}} \sigma(\bar{x})\sigma( \bar{y}),
\end{equation}
that is $\bar{H}_\mathit{\Delta}$ is the interaction energy in $\mathit{\Delta}$ with the removed
edge $\langle \bar{x}, \bar{y} \rangle$. Next, for
$\lambda \in \mathbb{R}$, we consider
\[
 P_{\mathit{\Delta}} (\lambda)= \int_{\mathcal{J}_q} \log \left\{ \int_{\mathbb{R}^{|\mathit{\Delta}|}} \exp\left[\lambda J_{\bar{x} \bar{y}} \sigma(\bar{x})\sigma( \bar{y}) - \bar{H}_\mathit{\Delta} (\sigma_{\mathit{\Delta}}|J,0) \right] \chi_{\mathit{\Delta}} (d \sigma_{\mathit{\Delta}}) \right\} \nu(d J).
\]
Then
\begin{equation}
 \label{acca}
 P_{\mathit{\Delta}} (1)= |\mathit{\Delta}| \int_{\mathcal{J}_q} p_{\mathit{\Delta}} (J,0) \nu (dJ).
\end{equation}
Clearly $P_{\mathit{\Delta}}$ is infinitely differentiable and
\begin{eqnarray*}
P'_{\mathit{\Delta}} (\lambda) & = &  \int_{\mathcal{J}_q} J_{\bar{x} \bar{y}} \langle \sigma(\bar{x})\sigma( \bar{y}) \rangle_{\bar{\pi}^\lambda_{\mathit{\Delta}}(J)}\nu(dJ), \\[.2cm]
 P''_{\mathit{\Delta}} (\lambda) & = &  \int_{\mathcal{J}_q}  \left(J_{\bar{x} \bar{y}} \right)^2 \left\{ \langle \left[\sigma(\bar{x})\sigma( \bar{y})\right]^2 \rangle_{\bar{\pi}^\lambda_{\mathit{\Delta}}(J)} - \langle \sigma(\bar{x})\sigma( \bar{y}) \rangle^2_{\bar{\pi}^\lambda_{\mathit{\Delta}}(J)}\right\} \nu(dJ),
\end{eqnarray*}
where the expectation $\langle \cdot \rangle_{\bar{\pi}^\lambda_{\mathit{\Delta}}(J)}$ is taken with respect to the measure (\ref{j5}) in which the interaction energy
is $-\lambda J_{\bar{x} \bar{y}} \sigma(\bar{x})\sigma( \bar{y}) + \bar{H}_\mathit{\Delta} (\sigma_{\mathit{\Delta}}|J,0)$.
For $\lambda=0$, this energy is independent of $J_{\bar{x} \bar{y}}$. Since $\nu$ is a product measure, by (\ref{pres4}) we have that $P'_{\mathit{\Delta}}(0) = 0$, whereas $P''_{\mathit{\Delta}} (\lambda)\geq 0$ for all $\lambda$. Therefore, $P'_{\mathit{\Delta}} (\lambda) \geq 0$ for all $\lambda \geq 0$, and hence
\begin{equation}
\label{integ2}
 P_{\mathit{\Delta}} (1) \geq P_{\mathit{\Delta}} (0),
\end{equation}
which is a kind of the first GKS inequality known for ferromagnets, see e.g. \cite{[A]}. For bounded interactions, a similar result was obtained in \cite{[Cont]} in Theorem 1. Then (\ref{integ2}) implies the superadditivity of $P_{\mathit{\Delta}}$, see Theorem 2 in \cite{[Cont]}, and thereby (\ref{pres5}), c.f. Corollary 2.1 in \cite{[Cont]} and Proposition 3.3.3 in \cite{[Bovier]}, page 37.

Now let us prove the second part of the theorem. Here we follow the proof of Theorem 3.10 in \cite{[KP]} or Theorem 5.1.3, page 268 in \cite{[A]}. For $t_1, t_2 \in \mathbb{R}$ and $\xi \in \mathcal{S}_p$, by the Jensen inequality we get from (\ref{j5b})
\begin{eqnarray*}
& & Z_{\mathit{\Delta}} (J, (t_1 + t_2) \xi) \geq Z_{\mathit{\Delta}} (J, t_1  \xi)\\[.2cm] & & \qquad \times\exp\left[ t_2 \sum_{x\in \mathit{\Delta}, \ y\in \mathit{\Delta}^c, \ x \sim y}J_{xy} \xi(y) \langle \sigma(x) \rangle_{\pi_{\mathit{\Delta}} (\cdot| J, t_1 \xi)}  \right].
\end{eqnarray*}
We set here first $t_1 = 0$, $t_2 = 1$, then $t_1 = - t_2 = 1$, and obtain
\begin{eqnarray*}
& & p_{\mathit{\Delta}}(J, 0) + \frac{1}{|\mathit{\Delta}|} \sum_{x\in \mathit{\Delta}, \ y\in \mathit{\Delta}^c, \ x \sim y}J_{xy} \xi(y) \langle \sigma(x) \rangle_{\pi_{\mathit{\Delta}} (\cdot| J, 0)}\\[.2cm] & & \qquad  \leq p_{\mathit{\Delta}}(J, \xi) \nonumber \\[.2cm] & & \leq p_{\mathit{\Delta}}(J, 0) + \frac{1}{|\mathit{\Delta}|} \sum_{x\in \mathit{\Delta}, \ y\in \mathit{\Delta}^c, \ x \sim y}J_{xy}  \langle \sigma(x) \sigma(y) \rangle_{\pi_{\mathit{\Delta}} (\cdot| J, \xi)}\nonumber
\end{eqnarray*}
Integrating this double inequality with respect to $\vartheta$ we get
\begin{eqnarray}
  \label{lipp}
& & \left\vert \int_{\mathcal{J}_q}\bar{p}_{\mathit{\Delta}} (J) \nu(dJ) - \int_{\mathcal{J}_q}{p}_{\mathit{\Delta}} (J,0) \nu(dJ) \right\vert \\[.2cm] & & \quad \leq   \frac{ 2 d |\partial \mathit{\Delta}|}{|\mathit{\Delta}|} \left( 2 c_\nu + a_\nu  \right), \nonumber
\end{eqnarray}
which clearly tends to zero along any van Hove sequence. Then (\ref{pres6}) follows from (\ref{pres5}). In getting (\ref{lipp}) we used the estimate (\ref{fz}) with $\varkappa =1$, as in (\ref{een}), and then (\ref{pres3}), (\ref{integ}), and (\ref{integG}).
$\square$

\section{The proof of the basic lemmas}
\label{5SE}

In this section, we assume that $J$, $\chi$, $q$, and $p$
are  as in Theorem \ref{1tm}. In the next lemma, which is a version of Lemma \ref{New1lm} for a one-point $\mathit{\Delta} = \{x\}$, we write
$\pi_x$ meaning $\pi_{\{x\}}$.
\begin{lemma}
  \label{1lm}
For every positive $\lambda$ and $\varkappa$, and for any $x\in \mathbb{Z}^d$ and $\xi\in \mathbb{R}^{\mathbb{Z}^d}$, one has
\begin{eqnarray}
  \label{j15}
& &  \int_{\mathbb{R}^{\mathbb{Z}^d}} \exp\left[ \lambda |\sigma(x)|^{p} \right]\pi_x (d \sigma |J,\xi) \\[.2cm] & & \quad \leq \exp\left[ C( \lambda , \varkappa) + 2 \varkappa \sum_{y \sim x} |\xi(y)|^{p} + 2 \varkappa^{1-q} \sum_{y\sim x}|J(x,y)|^q  \right], \nonumber  \end{eqnarray}
where
\begin{equation}
  \label{j16}
  C(\lambda , \varkappa) = \log C_{+} (\lambda + 2 d \varkappa) - \log C_{-} (2 d \varkappa).
\end{equation}
\end{lemma}
{\it Proof:} By (\ref{j4}) and (\ref{j5}) we have
\begin{eqnarray}
  \label{j17}
& &  \int_{\mathbb{R}^{\mathbb{Z}^d}} \exp\left[ \lambda |\sigma(x)|^{p} \right]\pi_x (d \sigma |J,\xi) \\[.2cm] & & \qquad  = \frac{1}{Z_x (J,\xi) }\int_{\mathbb{R}} \exp\left[ \lambda |u|^{p} + \sum_{y\sim x} J(x,y) u \xi(y) \right]\chi_x (d u), \nonumber
\end{eqnarray}
and
\begin{equation}
  \label{j18}
Z_x (J,\xi) =   \int_{\mathbb{R}} \exp\left[  \sum_{y\sim x} J(x,y) u \xi(y) \right]\chi_x (d u).
\end{equation}
By means of (\ref {fz}) with $p$ as in (\ref{j14a}), and by (\ref{lipq}), we get
\begin{eqnarray*}
 & & - \mathit{\Gamma} (\varkappa , q) \leq |J(x,y) u \xi(y)| \leq  \mathit{\Gamma} (\varkappa , q), \\[.2cm]
 & &  \mathit{\Gamma} (\varkappa , q) \ \stackrel{\rm def}{=} \ \varkappa\left( |u|^{p} + |\xi(y)|^{p}\right) + \varkappa^{1-q} |J(x,y)|^q.
\end{eqnarray*}
Applying these estimates in (\ref{j17}) (upper bound), and in (\ref{j18}) (lower bound), and taking into account (\ref{j13}) we readily get (\ref{j15}). $\square$ \vskip.1cm \noindent
The estimate just proven allows one to control the dependence of the integrals as in (\ref{j15}) on $\xi$ and $J$.
Note that the influence of $\xi$ is small for $\varkappa\ll \lambda$. However, for such $\varkappa$, the third term on the right-hand side of (\ref{j15}) is big, and vice versa.

Our next step is to extend the estimate (\ref{j15}) to arbitrary  $\mathit{\Delta} \Subset \mathbb{Z}^d$. For such a set
 $\mathit{\Delta}$, and for  $x\in \mathit{\Delta}$ and $\lambda >0$, we put
\begin{equation}
  \label{j19}
M_x (J, \lambda, \mathit{\Delta} |\xi) = \log \left\{ \int_{\mathbb{R}^{\mathbb{Z}^d}} \exp\left( \lambda |\sigma (x) |^{p}\right)\pi_\mathit{\Delta} (d\sigma|J,\xi)\right\}.
\end{equation}
To find an upper estimate for this function we integrate both sides of the estimate (\ref{j17}) with respect  to $\pi_{\mathit{\Delta}} ( d \sigma |J,\xi)$, which by (\ref{i2}) yields
\begin{eqnarray}
  \label{j20}
\exp\left[M_x (J, \lambda, \mathit{\Delta} |\xi) \right] & \leq & \exp\left( C(\lambda , \varkappa) + 2 \varkappa^{1-q} \sum_{y\sim x}|J(x,y)|^q \right. \\[.2cm] & + & \left. 2 \varkappa \sum_{y\sim x, \ y \in \mathit{\Delta}^c}  |\xi(y)|^{p}\right) \nonumber \\[.2cm] & \times & \int_{\mathbb{R}^{\mathbb{Z}^d}} \exp\left(2 \varkappa \sum_{y\sim x, \ y \in \mathit{\Delta}}  |\sigma(y)|^{p} \right) \pi_{\mathit{\Delta}} ( d \sigma |J,\xi) .\nonumber
\end{eqnarray}
Now we fix $\lambda>0$ and choose $\varkappa$ such that
\begin{equation}
  \label{j21}
  4 d w_0 \varkappa = \lambda/2.
\end{equation}
To estimate the integral in (\ref{j20}) we use the following
form of the H\"older inequality
\begin{equation}
  \label{j22}
  \int \left( \prod_{i=1}^n \varphi_i^{\alpha_i} \right) {\rm d}\mu
\leq \prod_{i=1}^n  \left(\int \varphi_i {\rm d}\mu
\right)^{\alpha_i},
\end{equation}
in which $\mu$ is a probability measure, $\varphi_i \geq 0$
(respectively, $\alpha_i \geq 0$), $i=1, \dots , n$, are integrable
functions (respectively, numbers such that $\sum_{i=1}^n \alpha_i
\leq 1$). Applying this inequality in (\ref{j20}) and taking into
account (\ref{j21}) we arrive at
\begin{eqnarray}
  \label{j23}
M_x (J, \lambda, \mathit{\Delta} |\xi) & \leq & C\left( \lambda , \frac{\lambda }{8 d w_0}\right) + 2 \left(\frac{\lambda}{8dw_0}\right)^{1-q} \sum_{y\sim x}|J_{x,y}|^q \\ & + & \frac{\lambda}{4d w_0} \sum_{y\sim x, \ y \in \mathit{\Delta}^c}  |\xi(y)|^{p} + \frac{1}{4dw_0} \sum_{y\sim x, \ y \in \mathit{\Delta}} M_y (J, \lambda, \mathit{\Delta} |\xi).  \nonumber
\end{eqnarray}
As the quantity we want to estimate appears in both sides of the latter inequality, we
obtain an upper bound for
\begin{equation}
  \label{j24}
 \|M (J, \lambda, \mathit{\Delta} |\xi)\|_w = \sum_{x\in \mathit{\Delta}} w(x) M_x (J, \lambda, \mathit{\Delta} |\xi) .
\end{equation}
\begin{lemma}
  \label{2lm}
Let $\xi$ be in $\mathcal{S}_p$. Then
\begin{eqnarray}
  \label{j25}
\|M (J, \lambda, \mathit{\Delta} |\xi)\|_w & \leq & 2|w| C\left(\lambda , \frac{ \lambda}{8dw_0}\right) \\[.2cm] & + & 4\left(\frac{\lambda}{8dw_0}\right)^{1-q} \|J\|^q_q  + \frac{\lambda}{2d}\sum_{x\in \mathit{\Delta}^c} |\xi(x)|^{p} w(x). \nonumber
\end{eqnarray}
\end{lemma}
{\it Proof:} We multiply both sides of (\ref{j23}) by $w(x)$, then sum up over $\mathit{\Delta}$, take into account (\ref{j8}) and (\ref{j9}), and obtain (\ref{j25})
. $\square$ \vskip.2cm \noindent
{\it Proof of Lemma \ref{New1lm}:}
By (\ref{j5}) and (\ref{j10}), for any $\delta >0$, we have
\begin{eqnarray}
  \label{j29}
& & \int_{\mathbb{R}^{\mathbb{Z}^d}} \exp\left( \lambda \|\sigma\|_{p}^{p} \right) \pi_{\mathit{\Delta}}( d\sigma|J,\xi) = \exp\left( \lambda \sum_{x\in \mathit{\Delta}^c} |\xi(x)|^{p} w(x)\right)\qquad  \\[.2cm]
& & \qquad  \quad \times \int_{\mathbb{R}^{\mathbb{Z}^d}} \prod_{x\in \mathit{\Delta}}\left[\exp\left(\delta |\sigma(x)|^{p}\right)  \right]^{\lambda w(x) /\delta}\pi_{\mathit{\Delta}}( d\sigma|J,\xi).  \nonumber
\end{eqnarray}
Take $\delta=  \lambda |w|$, so that
\[ \frac{\lambda}{\delta} \sum_{x\in \mathit{\Delta}} w(x) \leq 1,
\]
and apply in the last line the H\"older inequality (\ref{j22}). This yields, see (\ref{j19}) and (\ref{j24}),
\begin{eqnarray}
  \label{j30}
& &  \int_{\mathbb{R}^{\mathbb{Z}^d}} \exp\left( \lambda \|\sigma\|_{p}^{p} \right) \pi_{\mathit{\Delta}}(d\sigma|J,\xi) \\[.2cm]& & \quad \leq \exp\left( \lambda \sum_{x\in \mathit{\Delta}^c} |\xi(x)|^{p} w(x)\right)\exp\left( \frac{1}{ |w|} \|M(J, \lambda |w|, \mathit{\Delta}|\xi)\|_w \right). \nonumber
\end{eqnarray}
Now we apply the estimate (\ref{j25}) and obtain (\ref{New1}) with
\begin{eqnarray} \label{lipq1}
\mathit{\Upsilon}_1 (\lambda) & = & {2}C\left( \lambda |w|, \frac{\lambda|w|}{8dw_0}\right), \quad
\mathit{\Upsilon}_2 (\lambda) = \frac{4}{|w|^q}  \left(\frac{\lambda}{8dw_0}\right)^{1-q},\qquad \\[.2cm]
& & \qquad \quad \mathit{\Upsilon}_3 (\lambda) = \lambda (1 + 1/2d),  \nonumber
\end{eqnarray}
which completes the proof.
 $\square$. \vskip.1cm \noindent
As a corollary of (\ref{New1}) we get that
\begin{equation}
  \label{j28}
\limsup_{\mathit{\Delta}\nearrow \mathbb{Z}^d}\int_{\mathbb{R}^{\mathbb{Z}^d}} \exp\left( \lambda \|\sigma\|_{p}^{p} \right) \pi_{\mathit{\Delta}}( d\sigma|J,\xi) \leq \exp\left( \mathit{\Upsilon}_1(\lambda) + \mathit{\Upsilon}_2(\lambda) \|J\|_q^q  \right).
\end{equation}
{\it Proof of Lemma \ref{New2lm}:} We fix $f$, $\mathit{\Delta}$, $R$, $\xi$, and consider the function, c.f. (\ref{j4}) and (\ref{j5}),
\begin{eqnarray}
 \label{fz1}
& & \psi(J)  =  \int_{\mathbb{R}^{\mathbb{Z}^d}} f(\sigma) \pi_{\mathit{\Delta} } (d \sigma |J, \xi)  =
\frac{1}{Z_{\mathit{\Delta}} (J, \xi)} \int_{\mathbb{R}^{|\mathit{\Delta}|}} f(\sigma_{\mathit{\Delta}}  \times \xi_{\mathit{\Delta}^c}) \\[.2cm] & & \times \exp\left(\sum_{\langle x, y \rangle \in {\sf E}_{\mathit{\Delta}} } J_{xy} \sigma (x) \sigma (y) + \sum_{x \in \mathit{\Delta}, \ y\in \mathit{\Delta}^c, \ x \sim y} J_{xy} \sigma (x)  \xi (y) \right)\chi_{\mathit{\Delta}} (d \sigma_\mathit{\Delta}).
\nonumber
\end{eqnarray}
It is independent of $J_{xy}$ with both $x$ and $y$ in $\mathit{\Delta}^c$, and is everywhere differentiable
with respect to any $J_{xy}$. Set
\begin{equation}
 \label{fz2}
\mathit{\Psi}_{xy} (R) = \sup_{J \in B_q(R)} \left\vert \frac{\partial \psi(J) }{\partial J_{xy}}\right\vert.
\end{equation}
Then, for $J, J' \in B_q(R)$, we have
\begin{eqnarray}
 \label{fz3}
\left\vert \psi(J) - \psi(J') \right\vert & \leq & \sum_{\langle x, y \rangle \in {\sf E}_{\mathit{\Delta}} }
\left\vert J_{xy} - J'_{xy} \right\vert  \mathit{\Psi}_{xy} (R) \\[.2cm] & + & \sum_{x \in \mathit{\Delta}, \ y\in \mathit{\Delta}^c, \ x \sim y}\left\vert J_{xy} - J'_{xy} \right\vert  \mathit{\Psi}_{xy} (R). \nonumber
\end{eqnarray}
By (\ref{j12}), it follows that
\begin{equation}
 \label{fz4}
\left\vert J_{xy} - J'_{xy} \right\vert \leq \|J - J' \|_q \left[w(x) + w(y) \right]^{-1/q}.
\end{equation}
On the other hand, from (\ref{fz1}) we get
\begin{equation} \label{fz5}
\mathit{\Psi}_{xy} (R) \leq 2 \|f\|_{\infty} \sup_{J \in B_q(R)} \int_{\mathbb{R}^{\mathbb{Z}^d}} \left\vert \sigma(x) \tilde{\sigma}(y)  \right\vert  \pi_{\mathit{\Delta} } (d \sigma |J, \xi),
\end{equation}
where
\[
\tilde{\sigma}(y) = {\sigma}(y) \quad {\rm for} \ y \in \mathit{\Delta}, \quad \tilde{\sigma}(y) = \xi(y) \quad {\rm for} \ y \in \mathit{\Delta}^c , \ y \sim x.
\]
Now we apply (\ref{fz}) with $a = \left[w(x) + w(y) \right]^{-1/q}$, $b = |\sigma(x)|$, and $c = |\tilde{\sigma}(y)|$, take into account (\ref{lipq}), and obtain
\begin{eqnarray*}
\left[w(x) + w(y) \right]^{-1/q} \left\vert \sigma(x) \tilde{\sigma}(y)  \right\vert & \leq &
\left[w(x) + w(y) \right]\left(\left\vert \sigma(x)  \right\vert^p  + \left\vert \tilde{\sigma}(y)  \right\vert^p   \right) \qquad \\[.2cm] & + & \left[w(x) + w(y) \right]^{-q}. \nonumber
\end{eqnarray*}
Then we use this estimate in (\ref{fz3}), take into account also (\ref{fz4}), (\ref{fz5}), and (\ref{New1}), and obtain  (\ref{New2}) with
\begin{eqnarray*}
\mathit{\Theta}_1 (\mathit{\Delta}, R) & = & 8 d (1+ w_0) \left( \mathit{\Upsilon}_1(\lambda) + \mathit{\Upsilon}_2(\lambda) R^p\right)/\lambda  \\[.2cm] & + & 2 \left( \sum_{\langle x, y \rangle \in {\sf E}_{\mathit{\Delta}} } \left[w(x) +  w(y) \right]^{-q} \right. \\[.2cm] & + & \left. \sum_{x \in \mathit{\Delta}, \ y\in \mathit{\Delta}^c, \ x \sim y} \left[w(x) + w(y) \right]^{-q} \right), \\[.3cm] \mathit{\Theta}_2 (\mathit{\Delta}, R) & = & 4 d (1+ w_0) \left( 1 + 2 \mathit{\Upsilon}_3(\lambda)/\lambda \right).
\end{eqnarray*}
Here $\lambda$ can be taken arbitrarily, e.g. $\lambda = 1$. $\square$ \vskip.2cm \noindent
{\it Proof of Lemma \ref{integrlm}:} Let $\mu(J)$ be any random Gibbs measure. Then for $\nu$ as in (\ref{pres3}) and $\vartheta$ as in (\ref{pres2}), by Jensen's inequality we obtain from (\ref{J14aa})
\begin{equation}
  \label{liq}
\int_{\mathbb{R}^{\mathbb{Z}^d}\times \mathcal{J}_q} \sum_{x\in \mathbb{Z}^d} |\sigma(x)|^p w(x) \vartheta (d\sigma , dJ) \leq \mathit{ \Upsilon}_1 (1) +  2 \mathit{ \Upsilon}_2 (1)|w| a_\nu,
\end{equation}
see also (\ref{j8}) and (\ref{j12}). Due to the uniform bound in (\ref{pres3}), the right-hand side of (\ref{liq}) does not depend on the choice of the weight
$w(x)$ provided we keep fixed $w_0$ and $|w|$, see (\ref{lipq1}).  Thus, we can choose the weight such that
 $w(x)=1$ and $w(y) \leq 1$ for all other $y\in \mathbb{Z}^d$. For example, $w(y) = \exp(- \alpha |x-y|)$, c.f. (\ref{exa}). Then (\ref{integG}) follows from (\ref{liq}) with $c_\nu = {\rm RHS}(\ref{liq})$. The proof of
(\ref{integ}) follows from the estimate (\ref{New1}) in the same way.$\square$
\vskip.6cm \noindent
\large{\bf Acknowledgement}
\vskip.2cm \noindent
The authors are grateful to Stas Molchanov and Michael R\"ockner for valuable discussions.
This work was financially supported by the DFG through SFB 701: ``Spektrale Strukturen
und Topologische Methoden in der Mathematik"  and through the
research project 436 POL 125/0-1, which is cordially acknowledged by the authors.

\end{document}